\documentclass[a4paper,11pt]{article}
\pdfoutput=1 

\usepackage{jheppub} 

\usepackage[T1]{fontenc} 

\usepackage{graphicx}
\usepackage{comment}

\usepackage{subcaption}

\newcommand{\bea}{\begin{eqnarray}}
\newcommand{\eea}{\end{eqnarray}}
\def\no{\nonumber}

\newcommand{\p}{\partial}

\newcommand{\tr}{\textrm{Tr}}

\newcommand{\be}{\begin{equation}}
\newcommand{\ee}{\end{equation}}
\newcommand{\beq}{\begin{eqnarray}}
\newcommand{\eeq}{\end{eqnarray}}

\title{\boldmath On the relation between Volkov-Akulov and special conformal supersymmetry: a D3-brane perspective}

\author[a]{F. Coomans,}
\author[b]{B. Van Pol}

\affiliation[a,b]{Instituut voor Theoretische Fysica, KU Leuven,\\ Celestijnenlaan 200D B-3001 Leuven, Belgium}

\emailAdd{frederik.coomans@fys.kuleuven.be}
\emailAdd{bert.vanpol@fys.kuleuven.be}

\abstract{We obtain the complete superconformal symmetry transformations on the worldvolume of a D3-brane in an $AdS_5\times S^5$ background by using a coset superspace approach. We show that in the large $R$-limit we recover all supersymmetries present on the worldvolume of a D3-brane in a Minkowski background, in particular the Volkov-Akulov supersymmetry. We conclude with a proposal for a scheme to construct higher derivative invariants in $D=4$, $\mathcal{N}=4$ settings.}

\begin{document}
\maketitle

\newpage
\flushbottom

\section{Introduction}
In \cite{Bergshoeff:2013pia} new ways for constructing supersymmetric higher derivative invariants were investigated in settings where there are no known off-shell formulations. In particular, the action and supersymmetry transformation rules of the $D=4$, $\mathcal{N}=4$ Maxwell multiplet were deformed with higher derivative terms. This was done in such a way that at each order of the deformation the theory has 16 deformed Maxwell multiplet supersymmetries and 16 Volkov-Akulov (VA) type non-linear supersymmetries. The results were obtained by studying the worldvolume theory of the gauge-fixed D3-superbrane in a 10-dimensional Minkowski background.\\
It is still an open question if superconformal higher-derivative invariants in $D=4$, $\mathcal{N}=4$ supergravity exist. Constructing these invariants by using superconformal methods \cite{VanProeyen:1983wk,Kaku:1977pa,Ferrara:1977ij,Kaku:1978nz,Kaku:1978ea,Freedman:2012zz}
requires a superconformal extension of the rigid supersymmetric deformed Maxwell multiplet of \cite{Bergshoeff:2013pia}. A lot can be learned about such an extension by studying a D3-brane in an $AdS_5\times S^5$ background since the corresponding worldvolume theory is a superconformal one \cite{Claus:1997cq,Claus:1998mw,Claus:1998ts}. In this paper we aim to investigate the relation between the superconformal symmetry group of the $AdS_5\times S^5$ background and the VA supersymmetry group of the Minkowski background. We do this by applying the coset formalism \cite{Claus:1998yw,Claus:1998ra,Kallosh:1998zx,Castellani:1991et}, and constructing the transformation rules for the worldvolume theory of a D3-brane in these backgrounds. We investigate whether there is a way to mimic the superconformal symmetry induced by the $AdS_5\times S^5$ background in the case of the Minkowski background such that it could be used in the superconformal approach for constructing higher derivative terms. We choose a gauge for the worldvolume $\kappa$-symmetry that allows us to establish contact with the results of \cite{Bergshoeff:2013pia}.\\
The paper is organized as follows. In section \ref{CosetSuperspaces} we recap the coset formalism \cite{Claus:1998yw,Claus:1998ra,Kallosh:1998zx,Castellani:1991et}. We apply it to both backgrounds, Minkowski and $AdS_5\times S^5$, to obtain the background super isometries. In section \ref{D3braneWV} we discuss the D3-brane worldvolume theory, in particular we look at the worldvolume symmetries, discuss how to gaugefix the local symmetries and discuss the effects of the different backgrounds. In section \ref{D3WVcomparison} we discuss the large $R$ limit that will allow us to compare the symmetry transformations of both backgrounds. In section \ref{Conclusions} we present our conclusions and discuss a possible avenue for constructing higher derivative invariants. In Appendix \ref{Clifford} we collect our conventions regarding Clifford matrices. In Appendix \ref{SUMN} we present the $SU(2,2|4)$ algebra as well as two ways of decomposing the $SO(2,4)$ subalgebra, the $AdS$ and conformal decompositions. Finally, in Appendix \ref{AdSxSCoset} we provide details for the construction of the $AdS_5\times S^5$ background as a coset space.\\
\newline
\textbf{Notational conventions}\\
We use the following conventions for indices
\bea
\bar{M} &\qquad& \text{label for the coset generators in $\bold{K}$} \nonumber\\
\bar{I} &\qquad& \text{label for the stability group generators in $\bold{H}$} \nonumber\\
\Lambda=\{\bar{M},\bar{I}\} &\qquad& \text{label for the generators of the superalgebra $\bold{G}=\bold{K}\oplus\bold{H}$} \nonumber
\eea
\bea
\mathcal{A} &\qquad& \text{label for the collection of bosonic generators in $\bold{G}$} \no\\
\tilde{a},\tilde{b},\tilde{c}= 0,\ldots,4 &\qquad & \text{$AdS_5$ tangent space indices}\nonumber\\
a,b,c=0,\ldots, 3 &\qquad& \text{part of the $AdS_5$ tangent space indices such that $\tilde{a}=\{a,4\}$ }\nonumber\\
a',b',c'= 5,\ldots,9 &\qquad & \text{$S^5$ tangent space indices}\nonumber\\
A,B,C= 0,\ldots,9 &\qquad & \text{10D tangent space indices such that $A=\{\tilde{a},a'\}$}\nonumber\\
\tilde{m},\tilde{n},\tilde{p}=0,\ldots,4 &\qquad& \text{5D spacetime indices, associated with the $AdS_5$ space}
\nonumber\\
m,n,p=0,\ldots, 3 &\qquad& \text{part of the 5D spacetime indices such that $\tilde{m}=\{m,4\}$}\nonumber\\
m',n',p'=5,\ldots,9 &\qquad& \text{5D spacetime indices, associated with the $S^5$ space}\nonumber\\
M,N,P=0,\ldots,9 &\qquad& \text{10D spacetime indices such that $\{M=\tilde{m},m'\}$}\nonumber\\
\alpha,\beta,\gamma= 1,\ldots,4 &\qquad & \text{$so(2,4)$ spinor indices projected on the righthanded chiral subspace ($AdS_5$)}\nonumber\\
i,j,k= 1,\ldots,4 &\qquad & \text{$so(6)$ spinor indices projected on the righthanded chiral subspace ($S^5$)}\nonumber\\
\hat{\alpha},\hat{\beta},\hat{\gamma}= 1,\ldots,32 &\qquad & \text{$D=10$ Majorana-Weyl spinor indices}\nonumber\\
I,J,K=1,\ldots 4 &\qquad& \text{$so(6)$ spinor indices} \nonumber\\
\mu=0,\ldots 3 &\qquad & \text{coordinate indices of the worldvolume of the D3-brane} \nonumber
\eea

\section{Coset superspaces}\label{CosetSuperspaces}
In this section we briefly recap the formalism of Cartan forms on coset superspaces \cite{Claus:1998yw,Claus:1998ra,Kallosh:1998zx,Castellani:1991et}. We then use this formalism to write down the superisometries of Minkowski superspace and $AdS_5\times S^5$ superspace.\\
We consider the coset manifold $G/H$, where $G$ is a supergroup and $H\subset G$ is a subgroup. Each coset is represented by a coset representative $\mathcal{G}(Z)$, labelled by super-coordinates $Z^{M}=\{X^{M},\theta^{\alpha}\}$. 
Left-invariant Cartan 1-forms are defined as
\beq\label{Cartan::DefCartan}
L(Z)\equiv \mathcal{G}(Z)^{-1} d \mathcal{G}(Z).
\eeq 
Since $L(Z)$ is a group element close to the identity it is a $\bold{G}$ valued super 1-form
\beq
L(Z) = L^{\Lambda}\bold{T}_{\Lambda}=dZ^M L_M^{\Lambda}\bold{T}_{\Lambda},
\eeq
where $\bold{T}_{\Lambda}$ are the generators of the superalgebra $\bold{G}$ associated to $G$.\\
We consider two decompositions which will be useful. First there is the coset decomposition of the algebra, defined by $\bold{G}=\bold{K}\oplus\bold{H}$ where $\bold{H}$ is the Lie-algebra associated with the stability group $H$ of $G$, $\bold{G}$ is the Lie-algebra of $G$, and $\bold{K}$ collects the coset generators. We introduce the split of labels $\Lambda=(\bar{M},\bar{I})$, where $\bar{M}$ are the directions in $\bold{K}$ and $\bar{I}$ are the directions in $\bold{H}$. The second decomposition that we consider is a boson-fermion split of the algebra $\bold{G}=\bold{B}\oplus\bold{F}$, where $\bold{B}$ contains the bosonic generators $B_{A}$ and $\bold{F}$ the fermionic generators $F_{\alpha}$, and define the split of a $\bold{G}$-valued object A as
\beq
A=A^{\Lambda}\bold{T}_{\Lambda}= A^{\bold{B}}+A^{\bold{F}}=A^{\mathcal{A}}\bold{T}_{\mathcal{A}} + A^{\alpha}\bold{F}_{\alpha}.
\eeq
For the coset representative we choose the parametrization $\mathcal{G}(Z)=g(X) e^{\Theta}$, where $g(X)$ represents the bosonic coset representative of the coset space and 
\be
\Theta = \Theta^{\alpha}\bold{F}_{\alpha}=\theta^{\dot{\alpha}}e_{\dot{\alpha}}^{\;\;\alpha}(X)\bold{F}_{\alpha},
\ee
where $e_{\dot{\alpha}}^{\;\;\alpha}(X)$ determines the choice of fermionic coordinates.\\ 
In \cite{Claus:1998yw,Claus:1998ra} the complete geometric superfields $L(Z)$ and Killing superfields $\Sigma(Z)$ for a generic maximally supersymmetric superspace were constructed independent of the choice of coordinates (to all orders in $\theta$), we repeat their results here. The Cartan 1-forms and the parameters $\Sigma$ (defining the superisometries) are split as follows
\bea\label{Cartan::HCovSuperFSplit}
L&=& E + \Omega = E^{\bar{M}}\mathbf{K}_{\bar{M}} + \Omega^{\bar{I}} \mathbf{H}_{\bar{I}},\nonumber\\
\Sigma &=& \hat{\Xi}+\hat{\Lambda}=\hat{\Xi}^{\bar{M}}\mathbf{K}_{\bar{M}} + \hat{\Lambda}^{\bar{I}} \mathbf{H}_{\bar{I}},
\eea
where the parameters $\{\hat{\Xi}^{\bar{M}},\hat{\Lambda}^{\bar{I}}\}$ are defined in terms of the superisometries $\{\Xi^{M}, \Lambda^{\bar{I}}\}$ as
\be
\hat{\Xi}^{\bar{M}} = \Xi^{M} E_{M}^{\;\;\bar{M}},\qquad \hat{\Lambda}^{\bar{I}}=\Lambda^{\bar{I}} + \Xi^{M}\Omega_{M}^{\;\;\bar{I}}.
\ee
We will be interested in maximally supersymmetric superspaces where $\bold{F}\subset\bold{K}$ or $\mathbf{F}\cap \mathbf{H}=0$. Both the Minkowski and $AdS_5\times S^5$ backgrounds fall in this category. The bosonic generators are split into $\bold{B}=\{\bold{P}_a, \bold{M}_i \}$, with $\bold{P}_a\in \bold{K}$ and $\bold{M}_i\in\bold{H}$. We also consider the gravitino $L_{0}^{\mathbf{F}}$ to be vanishing.\\
Splitting $\bar{M}$ into bosonic $a$ and fermionic $\alpha$, the supervielbein is given by \cite{Claus:1998yw,Claus:1998ra}
\beq\label{Cartan::Supervielbein}
E_{M}^{\;\;\bar{M}}= 
\left(\begin{matrix}
  e_{\mu}^{\;\;b}(X) & 0 \\
  0 & e_{\dot{\alpha}}^{\;\;\beta}(X) 
 \end{matrix}\right) 
\left(\begin{matrix}
  \delta_{b}^{\;\;a} + (UA\mathcal{Y})_b^{\;\;a} & (UAB)_{b}^{\;\;\alpha} \\
  (A\mathcal{Y})_{\beta}^{\;\;a} & (AB)_{\beta}^{\;\;\alpha}
 \end{matrix}\right),
\eeq
where
\bea
A_{\alpha}^{\;\;\beta}&=&2\left(\frac{\sinh^{2}\mathcal{M}/2}{\mathcal{M}^2}\right)_{\alpha}^{\;\;\beta},\qquad B_{\alpha}^{\;\;\beta}=\left(\mathcal{M}\coth\mathcal{M}/2\right)_{\alpha}^{\;\;\beta},\nonumber\\
\mathcal{Y}_{\alpha}^{\;\;a}&=&-\Theta^{\delta}f_{\delta\alpha}^{\;\;\;\;a},\qquad \mathcal{M}_{\alpha}^{\;\;\beta}=f_{\alpha\gamma}^{\;\;\;\;A}\Theta^{\gamma}\Theta^{\delta}f_{\delta A}^{\;\;\;\;\beta},
\eea
and $f_{\Lambda\Sigma}^{\;\;\Gamma}$ are the structure constants of the algebra $\bold{G}$. The $e_{\mu}^{\;\;a}$ form the vielbein of the bosonic space and $e_{\dot{\alpha}}^{\;\;\alpha}$ is the matrix introduced in the boson-fermion parametrization of the coset representative. The matrix $U_{a}^{\;\;\alpha}$ and $\Theta^{\alpha}$ depend on the spinorial gauge choice $e_{\dot{\alpha}}^{\;\;\beta}$
\beq
U_{a}^{\;\;\alpha}=e_a^{\;\;\mu}\left[\theta^{\dot{\alpha}}\partial_{\mu}e_{\dot{\alpha}}^{\;\;\alpha} + (L_{0}^A)_{\mu}\theta^{\dot{\beta}}e_{\dot{\beta}}^{\;\;\beta}f_{A\beta}^{\;\;\;\;\alpha}\right].
\eeq
The superisometries, $\Sigma(Z)=\mathcal{G}^{-1}(Z)\Upsilon_0\mathcal{G}(Z)$, in general are determined completely in terms of the $\theta=0$ Killing superfields $\Sigma_0^{\Lambda}$, which we denote here by
\beq\label{Cartan::Killingsuperfield}
\Sigma_0^{\Lambda} T_{\Lambda}= \tilde{\xi}^{a}\mathbf{P}_a +\tilde{\epsilon}^{\alpha}\mathbf{F}_{\alpha} + \tilde{l}^{i} \mathbf{M}_i,
\eeq
where
\beq
\tilde{\xi}^a=\xi^{\mu}e_{\mu}^{\;\;a},\qquad \tilde{\epsilon}^{\alpha}=\epsilon^{\dot{\alpha}}e_{\dot{\alpha}}^{\;\;\alpha},\qquad
\tilde{l}^i=l^i + \xi^{\mu}\omega_{\mu}^{\;\;i}.
\eeq
In terms of the structure constants of $\bold{G}$, one can show \cite{Claus:1998yw,Claus:1998ra} that the superisometries are 
\bea \label{Cartan::SupIsometriesComps}
\Xi^{\mu} &=& \xi^{\mu} + \tilde{\epsilon}^{\beta}(\mathcal{M}^{-1}\tanh \mathcal{M}/2)_{\beta}^{\;\;\alpha} \mathcal{Y}_{\alpha}^{\;\;a}e_{a}^{\mu},\nonumber\\
\Xi^{\dot{\alpha}} &=& \left(\Theta^{\beta}\tilde{\xi}^{a}f_{a\beta}^{\;\;\;\;\alpha} + \Theta^{\beta}\tilde{l}^i f_{i\beta}^{\;\;\;\;\alpha} - \xi^{a}U_{a}^{\;\;\alpha}\right)e_{\alpha}^{\;\;\dot{\alpha}}\nonumber\\
& &+\tilde{\epsilon}^{\beta}(\mathcal{M}\coth\mathcal{M})_{\beta}^{\;\;\alpha}e_{\alpha}^{\;\;\dot{\alpha}} - \tilde{\epsilon}^{\gamma}\left(\mathcal{M}^{-1}\tanh\mathcal{M}/2\right)_{\gamma}^{\;\;\beta}(\mathcal{Y}U)_{\beta}^{\;\;\alpha}e_{\alpha}^{\;\;\dot{\alpha}}.
\eea
The variations of the superspace coordinates are given by
\beq\label{Cartan::RepSymmetries}
\delta X^{\mu}=-\Xi^{\mu},\qquad \delta\theta^{\dot{\alpha}}=-\Xi^{\dot{\alpha}}.
\eeq
In the next subsections we will use equations (\ref{Cartan::Supervielbein}) and (\ref{Cartan::RepSymmetries}) to write down the supervielbein and superisometries of the Minkowski and $AdS_5\times S^5$ background superspaces.

\subsection{Flat superspace}
As a warm-up we derive the isometries and vielbein of the Minkowski background. We start from the super Poincar\'e group $G$. The algebra is given by
\bea
[M_{AB},M_{CD}] &=& \eta_{A[C}M_{D]B}-\eta_{B[C}M_{D]A},\nonumber\\
\ [P_A,M_{BC}] &=&\eta_{A[B}P_{C]},\nonumber\\
\ [M_{AB},Q_{\hat{\alpha}}]&=&-\frac{1}{4}(\Gamma_{AB}Q)_{\hat{\alpha}},\nonumber\\ 
\{Q_{\hat{\alpha}},Q_{\hat{\beta}}\}&=& (\Gamma^A)_{\hat{\alpha}\hat{\beta}}P_A.
\eea
We make the split
\beq
\mathbf{H}=\{M_{AB}\},\qquad\text{and}\qquad\mathbf{K}=\{P_A,Q_{\hat{\alpha}}\}.
\eeq
This means that the indices of the previous section are chosen to be $\Lambda=\{A,[AB],\hat{\alpha}\}$, $\mathcal{A}=\{A,[AB]\}$, $\bar{I}=\{[AB]\}$, $ \bar{M}=\{a,\hat{\alpha}\}$.
The spacetime fields are given by
\beq
e^A_{M}=\delta_{M}^A,\qquad \psi_{M}=0,\qquad \omega_{M}^{AB}=0,
\eeq
and the solutions to the spacetime Killing equations ($\theta=0$) are
\beq
\xi^{M}=a^{M}+\lambda_{(M)}^{MN}x_{N},\qquad \epsilon^{\hat{\alpha}}(x)=\varepsilon_{0}^{\hat{\alpha}},\qquad l^{AB}=\lambda_{(M)}^{MN}\delta_{M}^A\delta_{N}^B,
\eeq
where $a^{M}$, $\lambda^{MN}_{(M)}$ and $\varepsilon_{0}^{\hat{\alpha}}$ are constant parameters. The matrix $\mathcal{M}$ vanishes, and the matrix $e_{\dot{\hat{\alpha}}}^{\;\;\hat{\beta}}=\delta_{\dot{\hat{\alpha}}}^{\;\;\hat{\beta}}$. The supervielbein (\ref{Cartan::Supervielbein}) is then given by
\beq\label{Mink::SuperVielbein}
E^{\hat{\alpha}}=d\theta^{\hat{\alpha}},\qquad E^{A}=dx^A+\bar{\theta}\hat{\Gamma}^A d\theta,
\eeq
where we suppressed the spinor indices in $\bar{\theta}\hat{\Gamma}^A d\theta= \theta^{\alpha}(\hat{\Gamma}^A)_{\alpha}^{\;\;\beta} d\theta_{\beta}$.\\
Plugging everything in (\ref{Cartan::SupIsometriesComps}), we obtain the well-known superisometries
\bea\label{Mink::isometries}
\delta x^{M}=-\Xi^{M}=-a^{M}-\lambda_{(M)}^{MN}x_{N} -\frac{1}{2}(\bar{\varepsilon}_0\hat{\Gamma}^{M}\theta+\text{h.c.}),\nonumber\\
\delta\theta^{\hat{\alpha}}=-\Xi^{\hat{\alpha}}=-\varepsilon^{\hat{\alpha}}_0 -\frac{1}{4}\lambda_{(M)}^{MN}(\hat{\Gamma}_{MN}\theta)^{\hat{\alpha}}.
\eea
To facilitate things later, we introduce projectors $\mathcal{P}_{Q,S}=\tfrac{1}{2}(1\mp\gamma_5)\otimes I_8$ (this is similar to what we will do for the $AdS_5\times S^5$ case (see also Appendix \ref{SUMN})) such that
\beq
\mathcal{P}_{Q}\theta = \theta_{\alpha}^{i},\qquad\mathcal{P}_{S}\theta = \vartheta_{\alpha}^{i},
\eeq
and we make a similar split for $\varepsilon_0$ into $\epsilon^i_{\alpha}$ and $\eta^i_{\alpha}$ respectively. In terms of these refined variables, we have for the transformations
\bea
\delta x^{M}&=& -a^{M}-\lambda_{(M)}^{MN}x_{N} -\frac{1}{2}\left[(\bar{\epsilon}_i\gamma^{m}\theta^i+\bar{\eta}_i\gamma^{m}\vartheta^i)\delta_m^{M} + (\bar{\epsilon}_i\vartheta^{i}-\bar{\eta}_i\theta^i)\delta^{M}_4 + (\bar{\epsilon}_i\vartheta^{j}  \right. \nonumber\\
& & \qquad+\bar{\eta}_i\theta^j) \left(\gamma'^{m'}\right)_{j}^{\;\;i}\delta_{m'}^{M} +\text{h.c.}],\nonumber\\
\delta \theta^i &=&-\epsilon^i -\frac{1}{4}\lambda^{mn}_{(M)}\gamma_{mn}\theta^i -\frac{1}{4}\lambda^{m'n'}_{(M)}\left(\gamma'_{m'n'}\right)_{j}^{\;\;i}\theta^j +\frac{1}{2}\lambda^{m4}_{(M)} \gamma_{m}\vartheta^i  -\frac{1}{2}\lambda^{m'4}_{(M)}\left(\gamma'_{m'}\right)_{j}^{\;\;i}\theta^j \nonumber\\
& & \qquad-\frac{1}{4}\lambda^{mn'}_{(M)}\gamma_m\left(\gamma'_{n'}\right)_{j}^{\;\;i}\vartheta^j,\nonumber \\
\delta \vartheta^i &=&-\eta^i -\frac{1}{4}\lambda^{mn}_{(M)}\gamma_{mn}\vartheta^i -\frac{1}{4}\lambda^{m'n'}_{(M)}\left(\gamma'_{m'n'}\right)_{j}^{\;\;i}\vartheta^j -\frac{1}{2}\lambda^{m4}_{(M)}\gamma_{m}\theta^i +\frac{1}{2}\lambda^{m'4}_{(M)}\left(\gamma'_{m'}\right)_{j}^{\;\;i}\vartheta^j\nonumber\\
& & \qquad -\frac{1}{4}\lambda^{mn'}_{(M)}\gamma_m\left(\gamma'_{n'}\right)_{j}^{\;\;i}\theta^j.
\eea 
This form of the isometries will be used to compare with the large $R$ limit of the $AdS_5\times S^5$ isometries.  For the $AdS_5\times S^5$ background, however, there is no mixing between the first five and last five directions, this means that $\lambda^{mn'}_{(M)}=\lambda^{4n'}_{(M)}=0$. For this reason we will set these equal to zero from here on out.

\subsection{$AdS_5\times S^5$ superspace}
To construct this superspace we start from the superconformal group $G=SU(2,2|4)$, which has $SO(4,2)\times SO(6)$ as its bosonic subgroup. The superalgebra is presented in more detail in Appendix \ref{SUMN}. For this supercoset the stability group $H$ is the product group $SO(4,1)\times SO(5)$, which is purely bosonic. The  $30+32$ generators of $SU(2,2|4)\supset SO(4,2)\times SO(6)$ are decomposed into $5+5$ translations $\tilde{P}_{\tilde{m}}$ and $P'_{m'}$, $10+10$ Lorentz generators $\tilde{M}_{\tilde{m}\tilde{n}}$ and $M'_{m'n'}$, and $16+16$ supersymmetries $Q_{\alpha}^{\;\;i}$ and $S_{\alpha}^{\;\;i}$. This superspace has $(10|32)$ coordinates (5 coordinates $x^{\tilde{m}}=\{x^m, \rho\}$ of $AdS$, 5 coordinates $z^{m'}$ of the sphere and 32 fermionic coordinates $\theta_{\alpha}^{\;\;}$ and $\vartheta_{\alpha}^{\;\;i}$). We have made the split
\beq
\mathbf{H}=\{\tilde{M}_{\tilde{m}\tilde{n}},M'_{m'n'}\},\qquad \mathbf{K}=\{\tilde{P}_{\tilde{m}},P'_{m'},Q_{\alpha}^{\;\;i},S_{\alpha}^{\;\;i}\}.
\eeq
This supercoset is an example of a maximal supersymmetric coset, i.e. all fermionic generators are in $\mathbf{K}$. We refer to Appendix \ref{AdSxSCoset} for a detailed discussion of the construction of the bosonic part of this coset space. Appendix \ref{AdSxSCoset} also contains a discussion on the choice of fermionic coordinates $e_{\dot{\hat{\alpha}}}^{\;\;\hat{\beta}}$ and the choice we make is given in equation (\ref{AdSSphere::FermCoordChoice}). In this section we will combine the results of Appendix \ref{AdSxSCoset} and construct the supervielbein and superspace isometries (pushing the details of the coset construction to Appendix \ref{AdSxSCoset}).\\
The metric of $AdS_5\times S^5$ is given by the sum of (\ref{AdSxS::AdSmetric}) and (\ref{AdSxS::Spheremetric})
\be \label{AdSxS::Fullmetric}
ds^2 =\rho^2 dx^2 +\left(\frac{R}{\rho}\right)^2d\rho^2 +\frac{4R^2}{(1+z^2)^2}dz^2.
\ee
The supervielbein (\ref{Cartan::HCovSuperFSplit}) of the geometry
\beq
E=E^{\bar{M}}K_{\bar{M}}=E^m \tilde{P}_m+E^{\rho}\tilde{P}_{\rho}+E^{m'}P'_{m'} + (\bar{Q}_i E^{i}_{Q} + \text{h.c.}) + (\bar{S}_i E^{i}_{S} + \text{h.c.}),
\eeq
has components
\bea\label{AdSxS::supervielbein}
E^m &=& \rho \left[dx^n\left(\delta_n^{\;\;m} - \frac{1}{2}\left(\frac{R}{\rho}\right)^2 \bar{\vartheta}_i\gamma_n\vartheta^j\bar{\vartheta}_j\gamma^m\vartheta^i\right) +\left(\frac{1}{2}d\bar{\theta}_i\gamma^m\theta^i+\frac{1}{4}\bar{\theta}_id\vartheta^{j}\bar{\theta}_j\gamma^m\theta^i+\text{h.c.}\right)\right. \nonumber\\ 
&\qquad& +\left. \left(\frac{R}{\rho}\right)^2\left(\frac{1}{2}d\bar{\vartheta}_i\gamma^m\vartheta^i+\frac{1}{4}\bar{\vartheta}_id\theta^j\bar{\vartheta}_j\gamma^m\vartheta^i+\text{h.c.}\right) \right] + \mathcal{O}(\theta\wedge\vartheta), \nonumber\\
E^{\rho} &=& \frac{R}{\rho}\left[d\rho -\frac{1}{2}\left(d\bar{\theta}_i\vartheta^i -d\bar{\vartheta}_i\theta^i + \text{h.c.}\right)\rho\right] +\mathcal{O}(\theta\wedge\vartheta), \nonumber\\
E^{m'} &=& e^{m'} - \frac{R}{2}\left(d\bar{\theta}_i\vartheta^{j}+ d\bar{\vartheta}_i\theta^j+dx^m\bar{\vartheta}_i\gamma_m\vartheta^j +\text{h.c.}\right)\left(u\gamma'^{m'}u^{-1}\right)_j^{\;\;i} + \mathcal{O}(\theta\wedge\vartheta), \nonumber\\
E^{i}_Q &=& \rho^{1/2} \left[d\theta^j -dx^m\gamma_m\vartheta^j + \frac{1}{3}\theta^k\left(d\bar{\vartheta}_k\theta^j -\bar{\theta}_kd\vartheta^j\right)\right]u_j^{\;\;i}+\mathcal{O}(\theta\wedge\vartheta), \nonumber\\
E^{i}_S &=& \rho^{-1/2} \left[d\vartheta^j + \frac{1}{3}\left(2d\bar{\theta}_k\vartheta^j-\bar{\vartheta}_{k}d\theta^j\right)+ dx^m\vartheta^k\bar{\vartheta}_k\gamma_m\vartheta^j\right]u_j^{\;\;i}+\mathcal{O}(\theta\wedge\vartheta). 
\eea
Here $\mathcal{O}(\theta\wedge\vartheta)$ stands for terms containing both $\theta^i$ and $\vartheta^i$. We do not include these terms because they will drop out when we discuss the D3-brane embedding and gauge-fixing in section \ref{D3braneWV}, where our gauge choice will set $\theta^i=0$. We have left the coordinates of the sphere unspecified here. They are coded in the coset representative $u$ and given in Appendix \ref{SCoset}. \\
The superisometries for the various coordinates are
\bea\label{AdSxS::Isometries}
\delta x^m &=& -\xi^m_C(x) - \frac{1}{2}\left(\bar{\epsilon}_i(x)\gamma^m\theta^i+\text{h.c.}\right) - \frac{1}{4}\left(\bar{\eta}_i\theta^j\bar{\theta}_j\gamma^m\theta^i+\text{h.c.}\right)\nonumber\\
&\qquad& -\left(\frac{R}{\rho}\right)^2\left[\lambda_{(K)}^m+\frac{1}{2}\left(\bar{\eta}_i\gamma^m\vartheta^i +\text{h.c.}\right) +\frac{1}{4}\left(\bar{\epsilon}_i(x)\vartheta^{j}\bar{\vartheta}_j\gamma^m\vartheta^i +\text{h.c.}\right)\right]\nonumber\\
&\qquad&+\mathcal{O}(\theta\wedge\vartheta),\nonumber\\
\delta \rho &=& \Lambda_D(x)\rho - \frac{1}{2}\left(\bar{\epsilon}_i(x)\vartheta^i-\bar{\eta}_i\theta^i + \text{h.c.}\right)\rho + \mathcal{O}(\theta\wedge\vartheta)\nonumber\\
\delta z^{m'} &=& -\xi^{m'}(z) + \frac{(1-z^{2})}{4}\left(\bar{\epsilon}_{i}(x)\vartheta^j+\bar{\eta}_i\theta^j + \text{h.c.}\right)\left(u\gamma'^{m'}u^{-1}\right)_j^{\;\;i} + \mathcal{O}(\theta\wedge\vartheta),\nonumber
\eea
\bea
\delta\theta^i &=& -\epsilon^i(x) -\frac{1}{2}\Lambda_D(x)\theta^i-\frac{1}{4}\Lambda_M(x)\cdot\gamma\theta^i-\frac{1}{4}\theta^j\Lambda^{IJ}_{SO(6)}\left(\hat{\gamma}'_{IJ}\right)_j^{\;\;i}\nonumber\\
&\qquad&-\left(\frac{R}{\rho}\right)^2\left[\lambda^m_{(K)}+\frac{1}{2}\left(\bar{\eta}_j\gamma^m\vartheta^j+\text{h.c.}\right)+ \frac{1}{4}\left(\bar{\epsilon}_j(x)\vartheta^k\bar{\vartheta}_k\gamma^m\vartheta^j +\text{h.c.}\right)\right] \gamma_m\vartheta^i\nonumber\\
&\qquad& -\frac{2}{3}\theta^j\left(2\bar{\eta}_{j}\theta^i-\bar{\theta}_j\eta^i\right) + \mathcal{O}(\theta\wedge\vartheta),\nonumber\\
\delta\vartheta^i &=& -\eta^i +\lambda_{(K)}^m  \gamma_m \theta^i+\frac{1}{2}\Lambda_D(x)\vartheta^i - \frac{1}{4}\Lambda_M(x)\cdot\gamma\vartheta^i-\frac{1}{4}\vartheta^j\Lambda_{SO(6)}^{IJ}\left(\hat{\gamma}'_{IJ}\right)_j^{\;\;i}\nonumber\\
&\qquad& -\frac{2}{3} \vartheta^j \left(2\bar{\epsilon}_j(x)\vartheta^i-\bar{\vartheta}_j\epsilon^i(x)\right)
+\mathcal{O}(\theta\wedge\vartheta)
\eea
These $AdS_5\times S^5$ isometries have been written in terms of $x$-dependent combinations of the superconformal parameters $a^m$, $\lambda_{(M)}^{mn}$, $\lambda_{(K)}^m$ and $\lambda_D$ as defined in (\ref{AdSCoset::ConformalIsometries}). We have defined
\beq
\epsilon^i (x) = \epsilon î + x^m\gamma_m \eta^i,
\eeq
and the supersymmetries and special supersymmetries are parametrized by $\epsilon$ and $\eta$. $\Lambda_{SO(6)}^{IJ}$ are the parameters of the $SO(6)$ R-symmetry, $\xi^{m'}(z)$ is given in (\ref{SCoset::sphereisometries}), and $\hat{\gamma}'_{IJ}$ are elements of the 6-dimensional Clifford algebra, realizing the translation between $SO(6)$ and $SU(4)$,
\beq
\Lambda^{IJ}_{SO(6)}=\frac{1}{2}\Lambda_{SU(4)i}^{\qquad \;\;j} (\hat{\gamma}'^{IJ})_{j}^{\;\;i}.
\eeq

\section{D3-brane Worldvolume Theory} \label{D3braneWV}
The world-volume action of a generic super D3-brane probe consists of two parts \cite{Horowitz:1991cd,Bergshoeff:1996tu,Bergshoeff:1998ha}
\beq
S=S_{\text{DBI}} + S_{\text{WZ}}. \label{D3brane::braneaction}
\eeq 
The worldvolume $\mathcal{M}_{4}$ is parametrized by $4$ coordinates $\sigma^{\mu}$.\\
The first term of the action, $S_{\text{DBI}}$, can be written as 
\beq
S_{\text{DBI}}=-\frac{1}{\alpha^2}\int_{\mathcal{M}_4} d^4\sigma \sqrt{-\det(G_{\mu\nu}+\alpha\mathcal{F}_{\mu\nu})}\,,
\eeq
It contains the induced metric
\beq
G_{\mu\nu} = E_{\mu}^{a}\eta_{ab}E_{\nu}^b,\qquad E_{\mu}^a = \partial_{\mu}Z^M E_M^a,
\eeq
where $E_{\mu}^{a}$ is the pull-back of the background vielbein $E^{a}$ to the worldvolume. $\alpha^2$ corresponds to the inverse brane tension and $\mathcal{F}_{\mu\nu}$ depends on the other fields on the world-volume. The superspace coordinates are now fields on the worldvolume $Z^M=Z^M(\sigma)$. \\
The Wess-Zumino component is an integral over the worldvolume of an appropriate $4$-form $\mathcal{A}_{4}$. It can be written as a closed $5$-form over a $5$-dimensional manifold which has the worldvolume as its boundary, and whose leading term is the pull-back of the background $5$ superform. It contains further terms that describe interaction of the extra tensor fields with the background forms of lower order. \\
Both terms of the brane action are by construction (separately) invariant under the background superisometries. The background isometries are now symmetries acting on fields, i.e. they depend on the worldvolume coordinates $\sigma$ through $Z^M(\sigma)$. Upon fixing the embedding of the brane in the background, the rigid background isometries will be realized on the remaining world-volume fields.

\subsection{Local symmetries of the worldvolume actions}
The D3-brane actions not only have global symmetries due to the background isometries, they also come with local symmetries. The first set of local symmetries of this action are the world-volume diffeomorphisms. They act as Lie-derivatives on the fields
\beq
\delta_{\text{loc.diff.}}Z^M=\zeta^{\mu}(\sigma)\partial_{\mu}Z^M,\qquad \delta_{\text{loc.diff.}}\mathcal{F}_{\mu\nu}=\zeta^{\rho}(\sigma)\partial_{\rho}\mathcal{F}_{\mu\nu} - 2 \partial_{[\mu}\zeta^{\rho}(\sigma)\mathcal{F}_{\nu]\rho}.
\eeq
The second local symmetry is called $\kappa$-symmetry \cite{Bergshoeff:1996tu,Bergshoeff:1998ha}, which is a local fermionic symmetry. Its parameter is a $10$-dimensional spinor $\kappa$, depending on the worldvolume coordinates. The variations $\delta Z^M$ of the world-volume fields are defined in terms of the supervielbein by
\bea\label{D3brane::KappaOnVielbein}
\delta_{\kappa} E^a &\equiv& \left(\delta_{\kappa}Z^M\right) E_M^{\;\;a}=0\nonumber\\
\delta_{\kappa} E^{\alpha} &\equiv& \left(\delta_{\kappa}Z^M\right) E_M^{\;\;\alpha} =\kappa^{\beta}(\sigma)\left(1+\Gamma^{\mathcal{C}}\right)_{\beta}^{\;\;\alpha}.
\eea
The matrix $\Gamma$ appears here as its charge conjugate $\Gamma^{\mathcal{C}}$ and it is an element of the 10-dimensional Clifford algebra, satisfying $\Gamma^2=1$, $\tr\Gamma=0$. It is a combination of gamma matrices and depends on the worldvolume fields. For the probe $D3$-brane, $\Gamma$ is given by 
\beq
\Gamma=\left(\begin{matrix}
0 & \beta_{-}\\
-\beta_{+} & 0
\end{matrix}\right),
\eeq
with
\bea
\beta_{-}&=&\frac{1}{\sqrt{-\det(G_{\mu\nu}+\alpha\mathcal{F}_{\mu\nu})}}\left(\sum_{k=0}^{2}\frac{(-\alpha)^k}{2^k k!}\gamma^{\mu_1\nu_1\ldots \mu_k\nu_k}\mathcal{F}_{\mu_1\nu_1}\ldots\mathcal{F}_{\mu_k\nu_k}\right) \Gamma^{D_3},\nonumber\\
\beta_{+}&=&\frac{1}{\sqrt{-\det(G_{\mu\nu}+\alpha\mathcal{F}_{\mu\nu})}}\left(\sum_{k=0}^{2}\frac{\alpha^k}{2^k k!}\gamma^{\mu_1\nu_1\ldots \mu_k\nu_k}\mathcal{F}_{\mu_1\nu_1}\ldots\mathcal{F}_{\mu_k\nu_k}\right) \Gamma^{D_3},
\eea
where $\Gamma^{D_3}=\frac{1}{4!}\epsilon^{\mu\nu\rho\sigma}\gamma_{\mu\nu\rho\sigma}$ and $\gamma^{\mu}$ are the pullback of the 10-dimensional gamma matrices.\\
We can invert the relations (\ref{D3brane::KappaOnVielbein}) to $\delta_{\kappa}X^{\mu}$ and $\delta_{\kappa}\theta^{\alpha}$ by using the inverse vielbein \cite{Claus:1998yw,Claus:1998ra}
\bea\label{D3brane::KappaSymmOnCoords}
\delta_{\kappa} X^{\mu} &=& -\kappa^{\beta}\left(1+\Gamma^{\mathcal{C}}\right)_{\beta}^{\;\;\alpha} \left(\mathcal{M}^{-1}\tanh \frac{\mathcal{M}}{2} \Upsilon\right)_{\alpha}^{\;\;\;\;a}e_a^{\mu},\nonumber\\
\delta_{\kappa} \theta^{\dot{\alpha}} &=& \kappa^{\gamma}\left(1+\Gamma^{\mathcal{C}}\right)_{\gamma}^{\;\;\beta} \left(\mathcal{M}\sinh^{-1}\mathcal{M} + \mathcal{M}^{-1} \tanh \frac{\mathcal{M}}{2} \Upsilon U \right)_{\beta}^{\;\;\;\;\alpha} e_{\alpha}^{\;\;\dot{\alpha}}.
\eea
Comparing with (\ref{Cartan::SupIsometriesComps}) we see that they almost act as supersymmetries, the difference being in the higher order fermion terms in $\delta_{\kappa}\theta^{\dot{\alpha}}$.\\
The irreducible $\kappa$ symmetries are defined by the algebraic constraint
\beq\label{D3brane::KappaSymmRedToIrred}
(1-\Gamma_{\text{cl}})\kappa=0,
\eeq
where $\Gamma_{\text{cl}}$ is the value of $\Gamma$ at the classical value of the fields, compatible with the gauge fixing and brane wave equations. We can write the irreducible $\kappa$ symmetries as
\beq
\kappa_{+}\equiv (1+\Gamma)\kappa_{*},
\eeq
where $\kappa_{*}$ is a solution to equation (\ref{D3brane::KappaSymmRedToIrred}),  $(1-\Gamma)\kappa_{*}=0$.

\subsection{The static gauge and the $Q$-gauge}
The embedding of the brane in the background can be described by identifying some of the worldvolume coordinates with the spacetime coordinates of the background. This `gauge fixing' has to be admissible, which means that it has to be compatible with the equations of motion derived from the probe-brane action, the branewave equations. We will consider an infinite extended brane and will therefore take the \textit{static gauge}
\beq\label{D3brane::StaticGauge}
\sigma^{\mu}=\delta_m^{\mu}x^m,
\eeq
where $x^m$ are $4$ coordinates of the background geometry. This gauge will only yield a stable configuration in specific backgrounds \cite{Claus:1998mw}. Two examples are the flat background and the $AdS\times S$ background where the $x^m$ have to be the directions parallel to the boundary of $AdS$.
The full transformation of the fields $Z^M(\sigma)$ is
\beq
\delta Z^M = \zeta^{\mu}\partial_{\mu}Z^M + \delta_{\text{global}} Z^M + \delta_{\kappa}Z^M,
\eeq 
where $\delta_{\text{global}} Z^M$ are the transformations in (\ref{Mink::isometries}) or (\ref{AdSxS::Isometries}). In order to preserve the gauge choice (\ref{D3brane::StaticGauge}) we need to impose the condition $\delta x^m=0$, leading to a decomposition law for $\zeta^{\mu}$.\\
In fixing the $\kappa$-symmetry we will be guided by the effects of the $AdS_5\times S^5$ background. There are two natural ways to gauge-fix the $\kappa$-symmetry and get rid of half of the fermionic gauge-degrees of freedom on the worldvolume. We can either set $\vartheta^i=0$ ($S$-gauge) or we can set $\theta^i=0$ ($Q$-gauge). However, the $S$-gauge is not admissible for the infinite static branes in their own near-horizon geometry. The classical values of the fields in the static gauge are $x^m=\delta^m_{\mu}\sigma^{\mu}$, $\rho=\text{constant}$, $z^{m'}=\text{constant}$, $\theta^{i}=\vartheta^i=0$, $\mathcal{F}_{\mu\nu}=0$ leading to $\Gamma_{cl}=\hat{\gamma}_{ST}$, where this matrix $\hat{\gamma}_{ST}$ is precisely the one used in the projector to define $Q$ and $S$ supersymmetry (Appendix \ref{ConformalDecomp}). This means that a gauge-fixing
\beq
0=\vartheta^i= \frac{1}{2}\left(1-\hat{\gamma}_{ST}\right)\Theta^i,
\eeq
will not affect the irreducible $\kappa$ symmetry and is not admissible. Since we are interested in the $AdS_5\times S^5$ background, this leaves us with the 'natural' choice of the $Q$-gauge, $\theta^i=0$. Imposing this condition will leave us with a decomposition law for $\kappa_{+}$.

\subsection{D3-Brane Worldvolume in Minkowski Background}\label{D3WVMink}
We consider the embedding of a D3-brane in a Minkowski background. The $\kappa$-symmetry transformation rules (\ref{D3brane::KappaSymmOnCoords}) become
\bea
\delta_{\kappa}x^M &=&-\frac{1}{2}\left[\left(\bar{\kappa}_{+Qi}\gamma^{m}\theta^i+\bar{\kappa}_{+Si}\gamma^{m}\vartheta^i\right)\delta_m^{M} + (\bar{\kappa}_{+Qi}\vartheta^{i}-\bar{\kappa}_{+Si}\theta^i)\delta^{M}_4\right. \nonumber\\
& & \qquad \left.+  (\bar{\kappa}_{+Qi}\vartheta^{j}+\bar{\kappa}_{+Si}\theta^j) \left(\gamma'^{m'}\right)_{j}^{\;\;i}\delta_{m'}^{M} +\text{h.c.}\right],\nonumber
 \eea
\beq
\delta_{\kappa}\theta^i = \kappa_{+Q}^i,\qquad \delta_{\kappa}\vartheta=\kappa_{+S}^i,
\eeq
where we have introduced the projections $\mathcal{P}_{Q,S}\kappa_+ =\kappa_{+Q,S}$. \\
As discussed in the previous section, the condition $\delta x^m=0$, needed to preserve the static gauge, and the $Q$-gauge condition $\theta^i=0$ (fixing the kappa gauge) give us two decomposition laws (up to cubic fermion terms)
\begin{equation}
\kappa_{+Q}^{i}=\epsilon^i-\frac{1}{2}\lambda^{m4}\gamma_{m}\vartheta^i,
\end{equation}
and
\beq
\zeta^{\mu}(\sigma)=a^{\mu}+\lambda_{(M)}^{\mu N}x_{N} -\frac{1}{2}\left[\bar{\vartheta}_i\gamma^{\mu}\left(\eta^i -\beta_+\left(\epsilon^i-\frac{1}{2}\lambda^{n4}\gamma_{n}\vartheta^i\right)\right)+\text{h.c.}\right].
\eeq
The remaining fields then have as transformation laws
\bea\label{Mink::WVfieldtransf}
\delta x^4 &=& \xi^{\mu} \p_{\mu}x^{4} -\frac{1}{2}\left[\bar{\vartheta}_i\gamma^{\mu}\left(\eta^i -\beta_+\left(\epsilon^i-\frac{1}{2}\lambda^{n4}\gamma_{n}\vartheta^i\right)\right)+\text{h.c.}\right] \p_{\mu} x^4\nonumber\\
& &-\xi^{4}-\left[\bar{\vartheta}_i\left(\epsilon^i-\frac{1}{4}\lambda^{m4}\gamma_m\vartheta^i\right)+\text{h.c.}\right], \nonumber\\
\delta x^{m'}&=& \xi^{\mu} \p_{\mu}x^{m'} -\frac{1}{2}\left[\bar{\vartheta}_i\gamma^{\mu}\left(\eta^i -\beta_+\left(\epsilon^i-\frac{1}{2}\lambda^{n4}\gamma_{n}\vartheta^i\right)\right)+\text{h.c.}\right] \p_{\mu} x^{m'}\nonumber\\
& &-\xi^{m'}-\left[(\bar{\vartheta}_i\epsilon^j-\frac{1}{4}\lambda^{m4}\vartheta_i\gamma_m\vartheta^j)\left(\gamma^{m'}\right)_{j}^{\;\;i}+\text{h.c.}\right], \nonumber\\
\delta \vartheta^i&=& \xi^{\mu} \p_{\mu} \vartheta^i-\frac{1}{2}\left[\bar{\vartheta}_j\gamma^{\mu}\left(\eta^j -\beta_+\left(\epsilon^j-\frac{1}{2}\lambda^{n4}\gamma_{n}\vartheta^j\right)\right)+\text{h.c.}\right] \p_{\mu}\vartheta^i\nonumber\\
& & -\frac{1}{4}\lambda^{mn}_{(M)}\gamma_{mn}\vartheta^i -\frac{1}{4}\lambda^{m'n'}_{(M)}\left(\gamma'_{m'n'}\right)_{j}^{\;\;i}\vartheta^j  -\left[\eta^i +\beta_+\left(\epsilon^i-\frac{1}{2}\lambda^{m4}\gamma_{m}\vartheta^i\right)\right],
\eea
where we used that $\kappa_{+S}=-\beta_+\kappa_{+Q}$ and defined $\xi^{M}=a^{M}+\lambda_{(M)}^{MN}x_{N}$.

\subsection{D3-Brane Worldvolume in $AdS_5\times S^5$ Background}\label{D3WVAdSxS}
In this section we consider the D3-brane embedded in its own near-horizon background, $AdS_5\times S^5$. Embedding a $D3$-brane in this background, the background coordinates are promoted to worldvolume fields and their transformations under $\kappa$ symmetry (\ref{D3brane::KappaSymmOnCoords}) are given by
\bea\label{AdSxS::KappaTransfWVfields}
\delta_{\kappa}x^m &=& -\frac{1}{2}\left(\bar{\kappa}_{+Qi}\gamma^m\theta^i +\text{h.c.}\right) -\frac{1}{4}\left(\bar{\kappa}_{+Si}\theta^j\bar{\theta}_{j}\gamma^m\theta^i +\text{h.c.}\right)\nonumber\\
&\qquad& -\left(\frac{R}{\rho}\right)^2\left[\frac{1}{2}\left(\bar{\kappa}_{+Si}\gamma^m\vartheta^i + \text{h.c.}\right) + \frac{1}{4}\left(\bar{\kappa}_{+Qi}\vartheta^j\bar{\vartheta}_j\gamma^m\vartheta^i+\text{h.c.}\right)\right]
+\mathcal{O}\left(\theta\wedge\vartheta\right),\nonumber\\
\delta_{\kappa}\theta^i &=& \kappa_{+Q}^i -\left(\frac{R}{\rho}\right)^2\left[\frac{1}{2}\left(\bar{\kappa}_{+Sj}\gamma^m\vartheta^j +\text{h.c.}\right) +\frac{1}{4}\left( \bar{\kappa}_{+Qj}\vartheta^{k}\bar{\vartheta}_{k}\gamma^m\vartheta^{j} +\text{h.c.}\right)\right]\gamma_m \vartheta^i \nonumber\\
&\qquad& -\frac{1}{3}\theta^j\left(2\bar{\kappa}_{+Sj}\theta^{i}-\bar{\theta}_j\kappa_{+S}^{i}\right) +\mathcal{O}\left(\theta\wedge\vartheta\right),\nonumber\\
\delta_{\kappa}\vartheta^i &=& \kappa_{+S}^{i}-\frac{1}{3}\vartheta^{j} \left(2\bar{\kappa}_{+Qj}\vartheta^i-\bar{\vartheta}_j\kappa_{+Q}^i\right) +\mathcal{O}\left(\theta\wedge\vartheta\right),\nonumber\\
\delta_{\kappa}\rho&=&-\frac{1}{2}\left(\bar{\kappa}_{+Qi}\vartheta^i-\bar{\kappa}_{+Si}\theta^i\right)\rho+ \text{ h.c} +\mathcal{O}\left(\theta\wedge\vartheta\right),\nonumber\\
\delta_{\kappa} z^{m'}&=&\frac{(1-z^{2})}{4}\left(\bar{\kappa}_{+Qi}\vartheta^j+\bar{\kappa}_{+Si}\theta^j-\text{ h.c}\right) \left(u\gamma'^{m'}u^{-1}\right)_{j}^{\;\;i}+\mathcal{O}\left(\theta\wedge\vartheta\right).
\eea
Again, the conditions $\delta x^m=0$ and $\theta^i=0$ imply two decomposition laws (up to cubic fermion terms)
\bea
\zeta^{\mu}(\sigma)&=&\xi^{\mu}_C(\sigma)+\frac{1}{2}\left(\bar{\epsilon}_i(\sigma)\gamma^{\mu}\theta^i+\text{h.c.}\right)+\left(\frac{R}{\rho}\right)^2\left[\lambda_{(K)}^{\mu}+\frac{1}{2}\left(\bar{\eta}_i\gamma^{\mu}\vartheta^i +\text{h.c.}\right)\right]\nonumber\\
&\qquad&+\frac{1}{2}\left(\bar{\kappa}_{+Qi}\gamma^{\mu}\theta^i+\text{h.c.}\right)+\left(\frac{R}{\rho}\right)^2\left[\frac{1}{2}\left(\bar{\kappa}_{+Si}\gamma^{\mu}\vartheta^i + \text{h.c.}\right)\right] \nonumber \\
&\qquad&+\mathcal{O}\left(\theta\wedge\vartheta\right)\,,
\eea
and,
\begin{equation}
\kappa_{+Q}^i=\epsilon^i(x)+\left(\frac{R}{\rho}\right)^2\lambda^m_{(K)} \gamma_m\vartheta^i\,.
\end{equation}
The remaining worldvolume fields (apart from the worldvolume vector which we ignore in this paper) are then $\rho(\sigma),z^{m'}(\sigma)$ and $\vartheta(\sigma)$ and their transformation rules are the following (up to cubic fermion terms)
\bea\label{AdsxS::WVFieldTransf}
\delta\vartheta^i &=&\hat{\xi}^{\mu}_C(\sigma)\partial_{\mu}\vartheta^i- \frac{1}{4}\Lambda_M(\sigma)\cdot\gamma\vartheta^i+\frac{1}{2}\Lambda_D(\sigma)\vartheta^i-\frac{1}{4}\vartheta^j\Lambda_{SO(6)}^{IJ}\left(\hat{\gamma}'_{IJ}\right)_j^{\;\;i}\no\\
&\qquad&-\frac{\beta_+}{R}\left(\frac{R}{\rho}\right)^2\lambda^m_{(K)} \gamma_m\vartheta^i-\frac{\beta_+}{R}\epsilon^i(\sigma)-\eta^i\,, \nonumber\\
\delta\rho&=& \hat{\xi}^{\mu}_C(\sigma)\partial_{\mu}\rho +\Lambda_D(\sigma)\rho -\left[\bar{\vartheta}_i\left(\epsilon^i +\frac{1}{2}\left(\frac{R}{\rho}\right)^2\lambda^{m}_{(K)}\gamma_m\vartheta^i\right)+ \text{h.c}\right]  \rho\nonumber\\
& &+\frac{1}{2}\left(\frac{R}{\rho}\right)^2\left[\bar{\eta}_i\gamma^{\mu}\vartheta^i+\bar{\vartheta}_i\gamma^{\mu}\frac{\beta_+}{R}\left(\epsilon^i(\sigma)+\left(\frac{R}{\rho}\right)^2\lambda^m_{(K)} \gamma_m\vartheta^i\right) + \text{h.c.}\right]\partial_{\mu}\rho, \nonumber
\eeq
\bea\label{AdsxS::WVFieldTransf}
\delta z^{m'} &=& \hat{\xi}^{\mu}_C(\sigma)\partial_{\mu}z^{m'} -\xi^{m'}(z)\nonumber\\
& & +\frac{(1-z^{2})}{2}\left[\bar{\epsilon}_i(\sigma)\vartheta^j + \frac{1}{2}\left(\frac{R}{\rho}\right)^2\Lambda^m_K\bar{\vartheta}_i\gamma_m\vartheta^j +\text{h.c.}\right] \left(u\gamma'^{m'}u^{-1}\right)_j^{\;\;i}\nonumber \\
& &+\frac{1}{2}\left(\frac{R}{\rho}\right)^2\left[\bar{\eta}_i\gamma^{\mu}\vartheta^i+\bar{\vartheta}_i\gamma^{\mu}\frac{\beta_+}{R}\left(\epsilon^i(\sigma)+\left(\frac{R}{\rho}\right)^2\Lambda^m_K \gamma_m\vartheta^i\right) + \text{h.c.}\right]\partial_{\mu}z^{m'} 
\eea
where
\begin{equation}
\hat{\xi}^{\mu}_C(\sigma)\equiv\xi^{\mu}_C(\sigma)+\Bigl(\frac{R}{\rho}\Bigr)^2\lambda^{\mu}_{(K)}\,.
\end{equation}


\section{From $AdS_5\times S^5$ to Minkowski: The Large $R$ limit}\label{D3WVcomparison}
We want to compare the resulting worldvolume transformations of the two backgrounds discussed in the previous section. Our aim is to establish a relation between the symmetries in $AdS_5\times S^5$ background and the Volkov-Akulov supersymmetries in the Minkowski background of \cite{Bergshoeff:2013pia}. In order to make an identification, we need to take a suitable large $R$ limit of the $AdS_5\times S^5$ background. We start out with a discussion of the proper limit.\\
To take this limit, it is convenient to change (background) spacetime coordinates. We define
\beq\label{AdSxS::CoordRedef}
\rho=e^{r/R},\qquad z^{m'}=\frac{\tilde{z}^{m'}}{2R}.
\eeq
The metric (\ref{AdSxS::Fullmetric}) then becomes
\beq
ds^2=e^{2r/R}dx^{\mu}\eta_{\mu\nu}dx^{\nu} + e^{-2r/R}dr^2 + \frac{1}{(1+\tilde{z}^2/(4R^2))^2}d\tilde{z}^2\,,
\eeq
which becomes Minkowski space in the limit $R\rightarrow \infty$. We also need to change variables in the algebra. The algebra we have used to derive the transformation rules in the previous sections relied on the conformal decomposition (Appendix \ref{ConformalDecomp}) and the right variables for the algebra are the ones of the AdS decomposition (Appendix \ref{SUMN::AdSDecompositionSec}). Equation (\ref{SUMN::RelConfAndAdsDecomp}) gives the relation between the various decompositions and is to be used to obtain the right variables. In particular this means that we redefine variables
\beq\label{AdSxS::FlatSpacelimit}
\tilde{\vartheta}^i= R \vartheta^i,\qquad \tilde{\eta}^i= R \eta^i \qquad\text{and}\qquad \tilde{\kappa}^i_{+S} = R\kappa^i_{+S}.
\eeq
Applying these redefinitions and taking the large $R$ limit nicely reduces the $AdS_5\times S^5$ supervielbein and isometries, (\ref{AdSxS::supervielbein}) and (\ref{AdSxS::Isometries}), to their Minkowski space equivalents, (\ref{Mink::SuperVielbein}) and (\ref{Mink::isometries}) (modulo the spacetime mixing requirement $\lambda^{mn'}_{(M)}=\lambda^{4n'}_{(M)}=0$). We now apply this to the transformations of the worldvolume fields in the $AdS_5\times S^5$ background, (\ref{AdsxS::WVFieldTransf}), and take the limit $R\rightarrow \infty$ to obtain 
\bea \label{AdSxS::WVTransfLargeR}
\delta r &=&\xi^{\mu}\partial_{\mu} r -\frac{1}{2}\left[\bar{\tilde{\vartheta}}_i\gamma^{\mu}\left(\tilde{\eta}^i -\beta_+\left(\epsilon^i -\frac{1}{2}\tilde{A}^{mS}\gamma_{m} \tilde{\vartheta}^{i}\right)\right) + \text{h.c.}\right]\p_{\mu} r\nonumber\\
& & -\tilde{A}^{S}-\tilde{A}^{Sn}x_n - \left[\bar{\tilde{\vartheta}}^i\left(\epsilon^i -\frac{1}{4}\tilde{A}^{mS}\gamma_m\tilde{\vartheta}^i\right) +\text{h.c.}\right], \nonumber\\
\delta \tilde{z}^{m'} &=&\xi^{\mu}\partial_{\mu}\tilde{z}^{m'} - \frac{1}{2}\left[\bar{\tilde{\vartheta}}_i\gamma^{\mu}\left(\tilde{\eta}^i -\beta_+\left(\epsilon^i -\frac{1}{2}\tilde{A}^{mS}\gamma_{m} \tilde{\vartheta}^{i}\right)\right) + \text{h.c.}\right]\p_{\mu} \tilde{z}^{m'}\nonumber\\
& & -\tilde{\xi}^{m'}  - \left[\left(\bar{\epsilon}_i-\frac{1}{4}\tilde{A}^{mS}\bar{\tilde{\vartheta}}_i\gamma_m\right)\tilde{\vartheta}^j (\gamma'^{m'})_j^{\;\;i} +\text{h.c.}\right], \nonumber \\
\delta \tilde{\vartheta}^{i} &=& \xi^{\mu}\partial_{\mu} \tilde{\vartheta}^{i} -\frac{1}{4}A_{(M)}^{mn}\gamma_{mn}\tilde{\vartheta}^{i}\nonumber \\
& & -\frac{1}{4}\Lambda^{IJ}_{SO(6)}(\hat{\gamma}'_{IJ})^{\;\;i}_j \tilde{\vartheta}^{j} - \left[\tilde{\eta}^{i} + \beta_+ \left(\epsilon^i -\frac{1}{2}\tilde{A}^{mS}\gamma_{m} \tilde{\vartheta}^{i}\right)\right]
\eea
where $\xi^{\mu}=\tilde{A}^{\mu}+r\tilde{A}^{\mu S}+\tilde{A}^{\mu n}x_n$.\\
Making the identifications
\bea
x^5_{\text{Mink}}=r,\quad x^{m'}_{\text{Mink}}=\tilde{z}^{m'},\quad \vartheta^i_{\text{Mink}} = \tilde{\vartheta}^i\nonumber\\
\lambda^{m4}_{\text{Mink}}=\tilde{A}^{mS},\quad \eta^i_{\text{Mink}}=\tilde{\eta}^i,\quad \lambda^{mn}_{(M),\text{Mink}}=\tilde{A}^{mn}_{(M)},
\eea 
and
\be
\xi^4_{\text{Mink}}=\tilde{A}^S+\tilde{A}^{Sm}x_m,
\ee
where the subscript Mink refers to the quantities in (\ref{Mink::WVfieldtransf}),
we can compare (\ref{AdSxS::WVTransfLargeR}) with (\ref{Mink::WVfieldtransf}) and we find an exact match between the worldvolume transformation rules.\\
However, there seems to be no way to link Minkowski background symmetries to the $AdS_5\times S^5$ symmetries without introducing a length scale, not at all a surprising result. The reason for this is quite simple and can be found by looking at the conformal algebra. The algebra corresponding to our $AdS_5\times S^5$ space was given in (\ref{SUMN::ConfDecAlgebra}). We are interested in the anti-commutators of the fermionic generators which we repeat here for convenience
\bea
\{Q_{\alpha}^{\;\;i},\bar{Q}_{j}^{\;\;\beta} \} &=& \delta_j^{\;\;i} (\gamma^{a})_{\alpha}^{\;\;\beta}P_a,\qquad \{S_{\alpha}^{\;\;i},\bar{S}_{j}^{\;\;\beta} \} = \delta_j^{\;\;i} (\gamma^{a})_{\alpha}^{\;\;\beta}K_a,\nonumber\\
\{Q_{\alpha}^{\;\;i},\bar{S}_{j}^{\;\;\beta} \} &=& \delta_j^{\;\;i}\delta_{\alpha}^{\;\;\beta}D+  \delta_j^{\;\;i} (\gamma^{ab})_{\alpha}^{\;\;\beta}M_{ab} - 2 \delta_{\alpha}^{\;\;\beta}U_j^{\;\;i}.
\eea
Before we take the limit $R\rightarrow \infty$, we need to write these anticommutators in the notation of the $AdS$ decomposition. Using the relations in section \ref{ConformalDecomp} we find
\bea
\left\{\left(\mathcal{P}_Q \tilde{\mathcal{Q}}\right)_{\hat{\alpha}}^{\;\;i},\left(\overline{\mathcal{P}_Q \tilde{\mathcal{Q}}}\right)_{j}^{\;\;\hat{\beta}} \right\} &=& -\frac{1}{2}\delta_j^{\;\;i} (\mathcal{P}_Q\hat{\gamma}^{aT})_{\hat{\alpha}}^{\;\;\hat{\beta}}\left(\tilde{P}_a+\frac{2}{R}\tilde{M}_{aS}\right),\\
 \left\{\left(\mathcal{P}_S \tilde{\mathcal{Q}}\right)_{\hat{\alpha}}^{\;\;i},\left(\overline{\mathcal{P}_S \tilde{\mathcal{Q}}}\right)_{j}^{\;\;\hat{\beta}} \right\} &=& -\frac{1}{2}\delta_j^{\;\;i}(\mathcal{P}_S\hat{\gamma}^{aT})_{\hat{\alpha}}^{\;\;\hat{\beta}}\left(\tilde{P}_a-\frac{2}{R}\tilde{M}_{aS}\right),\nonumber\\
\left\{\left(\mathcal{P}_Q \tilde{\mathcal{Q}}\right)_{\hat{\alpha}}^{\;\;i},\left(\overline{\mathcal{P}_S \tilde{\mathcal{Q}}}\right)_{j}^{\;\;\hat{\beta}} \right\} &=& -\frac{1}{2}\delta_j^{\;\;i}\left(\mathcal{P}_Q \right)_{\hat{\alpha}}^{\;\;\hat{\beta}}\tilde{P}_S+\frac{1}{2R}\delta_j^{\;\;i} (\mathcal{P}_Q\hat{\gamma}^{ab})_{\hat{\alpha}}^{\;\;\hat{\beta}}M_{ab} - \frac{1}{R} \left(\mathcal{P}_Q\right)_{\hat{\alpha}}^{\;\;\hat{\beta}}U_j^{\;\;i}.\nonumber
\eea
From these commutation relations it is clear that in the limit $R\rightarrow \infty$ the right hand side of the first two commutators reduces to a translation. In other words, the distinction between the operator $P_a$ and the operator $K_a$ disappears. The conformal structure is an $\mathcal{O}\left(\frac{1}{R}\right)$-effect, and requires a length scale used for separation to work.\\
In light of this it is also clear why the major difference of the Volkov-Akulov supersymmetries of \cite{Bergshoeff:2013pia} and conformal supersymmetry rests in the Volkov-Akulov supersymmetries anti-commuting into translations. There simply is no length scale from the background available to make the distinction between translations and special conformal transformations. Let us look at the relation with the results from \cite{Bergshoeff:2013pia} a bit closer. In order to really compare with \cite{Bergshoeff:2013pia}, we should write our transformations in a form that looks like (only considering the fermionic symmetries now)
\begin{align}
\delta \phi^I &\sim \left(\bar{\lambda}_i\Gamma^{\mu} \epsilon^{2i} +\bar{\lambda}_i\Gamma^{\mu} \beta_+\epsilon^{1i}\right)\p_{\mu}\phi^I, \no \\
\delta \lambda^{i} &\sim \epsilon^{2i} + \beta_+\epsilon^{1i}.
\end{align}
Looking at the transformations (\ref{AdSxS::WVTransfLargeR}), we find
\be
\epsilon^1 = \epsilon^i -\frac{1}{2}\lambda^{n4}\gamma_n \vartheta^i, \quad \epsilon^2 = \eta^i +2\epsilon^i -\frac{1}{2}\lambda^{n4}\gamma_n \vartheta^i.
\ee
In order to make the appearance of the Volkov-Akulov symmetry apparent, we define the parameters
\begin{align}
\varepsilon&=\epsilon^1 = \epsilon^i -\frac{1}{2}\lambda^{n4}\gamma_n \vartheta^i,\no \\
\zeta&=\epsilon^2-\epsilon^1 = \eta^i +\epsilon^i,
\end{align}
suggesting that the generators for supersymmetry and Volkov-Akulov symmetry will be
\be
\left(Q_{\text{SUSY}}\right)_{\hat{\alpha}}^{\;\;i}=\left(\mathcal{P}_Q \tilde{\mathcal{Q}}\right)_{\hat{\alpha}}^{\;\;i},\qquad \left(Q_{\text{VA}}\right)_{\hat{\alpha}}^{\;\;i}=\left(\mathcal{P}_Q \tilde{\mathcal{Q}}\right)_{\hat{\alpha}}^{\;\;i} + \left(\mathcal{P}_S \tilde{\mathcal{Q}}\right)_{\hat{\alpha}}^{\;\;i}.
\ee The corresponding algebra becomes
\bea
\left\{\left(Q_{\text{SUSY}}\right)_{\hat{\alpha}}^{\;\;i},\left(\overline{Q_{\text{SUSY}}}\right)_{j}^{\;\;\hat{\beta}} \right\} &=& -\frac{1}{2}\delta_j^{\;\;i} (\mathcal{P}_Q\hat{\gamma}^{aT})_{\hat{\alpha}}^{\;\;\hat{\beta}}\tilde{P}_a,\\ 
\left\{\left(Q_{\text{VA}}\right)_{\hat{\alpha}}^{\;\;i},\left(\overline{Q_{\text{VA}}}\right)_{j}^{\;\;\hat{\beta}} \right\} &=& -\frac{1}{2}\delta_j^{\;\;i}(\hat{\gamma}^{aT})_{\hat{\alpha}}^{\;\;\hat{\beta}}\tilde{P}_a -\frac{1}{2}\delta_j^{\;\;i}\left(\delta \right)_{\hat{\alpha}}^{\;\;\hat{\beta}}\tilde{P}_S - \frac{1}{R} \left(\delta\right)_{\hat{\alpha}}^{\;\;\hat{\beta}}U_j^{\;\;i},\nonumber\\
\left\{\left(Q_{\text{SUSY}}\right)_{\hat{\alpha}}^{\;\;i},\left(\overline{Q_{\text{VA}}}\right)_{j}^{\;\;\hat{\beta}} \right\} &=& -\frac{1}{2}\delta_j^{\;\;i} (\mathcal{P}_Q\hat{\gamma}^{aT})_{\hat{\alpha}}^{\;\;\hat{\beta}}\tilde{P}_a -\frac{1}{2}\delta_j^{\;\;i}\left(\mathcal{P}_Q \right)_{\hat{\alpha}}^{\;\;\hat{\beta}}\tilde{P}_S - \frac{1}{R} \left(\mathcal{P}_Q\right)_{\hat{\alpha}}^{\;\;\hat{\beta}}U_j^{\;\;i},\nonumber
\eea
where we clearly see the appearance of translations and shift-symmetries of the scalar fields in the anti-commutators of the Volkov-Akulov-symmetry.


\section{Conclusions}\label{Conclusions}
We compared the worldvolume transformation rules of a D3-brane embedded in a Minkowski background with those of a D3-brane embedded in an $AdS_5\times S^5$ background. We obtained a relation between the special supersymmetry transformations induced by the $AdS_5\times S^5$ background and the Volkov-Akulov symmetries related to the Minkowski background. In order to relate one to the other, one needs to introduce a length scale, a result that is reaffirmed by looking at the algebra. The existence of a length scale in the algebra associated to the $AdS_5\times S^5$ background allows for the distinction between translations and special conformal translations as an $\mathcal{O}\left(\frac{1}{R}\right)$-effect. When this effect is very small (at large $R$) this distinction disappears, and it is therefore no surprise that in a Minkowski background one only finds supersymmetry transformations that anti-commute into translations and shift-symmetries (i.e. the $16$ supersymmetries $+$ $16$ Volkov-Akulov symmetries of \cite{Bergshoeff:2013pia}).\\
The question remains then whether we can construct higher derivative invariants coupled to supergravity in the $D=4$, $\mathcal{N}=4$ setting with VA-type symmetries. We will provide a tentative scheme for constructing these higher derivative invariants. Having established a relation between the conformal symmetry inherited by the $AdS_5\times S^5$ background and the Volkov-Akulov symmetry due to the Minkowski background, we can use this relation as a tool for the construction of higher derivative invariants. The idea is to perform a construction of higher derivative invariants using superconformal methods in the theory of the brane embedded in $AdS_5\times S^5$, followed by making the redefinitions (\ref{AdSxS::CoordRedef}) and (\ref{AdSxS::FlatSpacelimit}), and then taking the limit necessary to obtain the Minkowski background. Our gauge choice to fix $\kappa$-symmetry is special in the sense that it has an easy limit to obtain the worldvolume theory of a D3-brane in a Minkowski background. However, for the practical application of superconformal methods it is less convenient since it does not have a simple, linearly realised form for the supersymmetries. How to fix the $\kappa$-symmetry gauge in the $AdS_5\times S^5$ background in the most natural way is still an open problem \cite{Metsaev:1998hf}. The gauge choice we made in this paper, however, makes the relation with the transformation rules in the Minkowski background clear, and should be related to this unknown gauge choice by field redefinitions. If we can find such a gauge choice to simplify the construction of higher derivative invariants, we can modify the scheme by starting from this case with (as of yet) unknown $\kappa$-symmetry gauge to construct higher derivative invariants using superconformal methods. Once these are constructed field redefinitions will transform these higher derivative invariants to the gauge used in this paper, the $Q$-gauge. It is then only a matter of taking the large $R$-limit to obtain higher derivative invariants in the desired $D=4$, $\mathcal{N}=4$ setting with VA-type symmetries.

\acknowledgments
We thank A. Van Proeyen for useful conversations. The work of B.V.P was supported in part by the FWO - Vlaanderen, Project No. G.0651.11, and in part by the Interuniversity Attraction Poles Programme
initiated by the Belgian Science Policy (P7/37).

\appendix
\section{Clifford algebras}\label{Clifford}
We follow the notation and conventions of \cite{Freedman:2012zz}. For ease of use, we collect here a few of the Clifford algebra's and their realisations. For the two background cases we need Clifford matrices tailored to their needs. We start of by constructing the algebras needed for the $AdS_5\times S^5$ case ($SO(2,4)$ and $SO(6)$) and conclude with a compatible $SO(1,9)$ construction for the Minkowski background.

\subsection{The $SO(2,4)$ Clifford algebra}
We extend the $SO(1,3)$ Clifford matrices by two more matrices as follows
\beq
\hat{\Gamma}_a=\gamma_a\otimes\sigma_1,\qquad \hat{\Gamma}_S=\gamma_4\otimes\sigma_1,\qquad \hat{\Gamma}_T=I_4\otimes(-i\sigma_2).
\eeq
The $\gamma_a$ are the $SO(1,3)$ gamma matrices and $\gamma_{4} =-i\gamma_0\gamma_1\gamma_2\gamma_3$. We define
\beq
\hat{\Gamma}_{*}=-i\hat{\Gamma}_0\hat{\Gamma}_1\ldots\hat{\Gamma}_S\hat{\Gamma}_T =I\otimes\sigma_3.
\eeq
Since we are in $6$ dimensions the minimal spinor is a Weyl spinor, the conformal spinor in $4$ dimensions. We will restrict to righthanded chiral spinors, $\hat{\Gamma}_*\lambda=-\lambda$. We restrict $\hat{\Gamma}_{ab}$ to the righthanded chiral subspace\footnote{This means that $\hat{\gamma}_{\tilde{a}\tilde{b}}=\hat{\Gamma}_{\tilde{a}\tilde{b}}\tfrac{1}{2}\left(1-\hat{\Gamma}_{*}\right)$.} ($\hat{\Gamma}\rightarrow\hat{\gamma})$
\beq\label{Clifford::24RHproj}
\hat{\gamma}_{ab}=\gamma_{ab},\qquad \hat{\gamma}_{aS}=\gamma_a\gamma_5,\qquad \hat{\gamma}_{aT}=-\gamma_a,\qquad \hat{\gamma}_{ST}=-\gamma_5.
\eeq
These matrices satisfy the relations
\beq\label{Clifford::ThatOneRel}
(\hat{\gamma}_{\hat{a}\hat{b}})_{\hat{\alpha}}^{\;\;\hat{\beta}}(\hat{\gamma}^{\hat{a}\hat{b}})_{\hat{\gamma}}^{\;\;\hat{\delta}} =2\delta_{\hat{\alpha}}^{\;\;\hat{\beta}}\delta_{\hat{\gamma}}^{\;\;\hat{\delta}} - 8\delta_{\hat{\alpha}}^{\;\;\hat{\delta}}\delta_{\hat{\gamma}}^{\;\;\hat{\beta}},\qquad 
(\hat{\gamma}_{\hat{a}\hat{b}})_{\hat{\alpha}}^{\;\;\hat{\beta}}(\hat{\gamma}^{\hat{c}\hat{d}})_{\hat{\beta}}^{\;\;\hat{\alpha}} =-8\delta_{[\hat{a}}^{\;\;\hat{c}}\delta_{\hat{b}]}^{\;\;\hat{d}},
\eeq
where for this section $\hat{\alpha}=1,\ldots 8$ and $\hat{a}=\{a,S,T\}$.

\subsection{The $SO(6)$ Clifford algebra}
We extend the $SO(5)$ Clifford matrices by one more matrix
\beq
\hat{\Gamma}^{'}_{a'}=\gamma^{'}_{a'}\otimes\sigma_2,\qquad \hat{\Gamma}^{'}_{S'}=\gamma^{'}_{9}\otimes\sigma_2,\qquad \hat{\Gamma}^{'}_{T'}=I_4\otimes\sigma_1,
\eeq
where $\gamma^{'}_{a'}$ are the $SO(4)$ gamma matrices and $\gamma^{'}_{9}$ is given by $\gamma^{'}_{9}=-\gamma^{'}_{5}\gamma^{'}_{6}\gamma^{'}_{7}\gamma^{'}_{8}$. We define
\beq
\hat{\Gamma}^{'}_{*}=-i\hat{\Gamma}^{'}_{5}\hat{\Gamma}^{'}_{6}\ldots\hat{\Gamma}^{'}_{S'} \hat{\Gamma}^{'}_{T'} =I\otimes\sigma_3.
\eeq
Like before we will restrict to righthanded chiral spinors, and identify
\beq\label{Clifford::06RHproj}
\hat{\gamma}^{'}_{a'b'}=\gamma^{'}_{a'b'},\qquad \hat{\gamma}^{'}_{a'S'}=\gamma^{'}_{a'}\gamma^{'}_{9},\qquad \hat{\gamma}^{'}_{a'T'}=i\gamma^{'}_{a'},\qquad \hat{\gamma}^{'}_{S'T'}=i\gamma^{'}_{9}.
\eeq
These matrices satisfy a similar relation as (\ref{Clifford::ThatOneRel})

\subsection{The $SO(1,9)$ Clifford algebra}
We will use a decomposition of $10$-dimensional $\gamma$-matrices $\hat{\Gamma}^{\text{(10D)}}_{M}$ into the $SO(1,4)$ and $SO(5)$ matrices as follows
\bea
\hat{\Gamma}^{\text{(10D)}}_{\tilde{m}}=\gamma_{\tilde{m}}\otimes I_4 \otimes \sigma_1,&\qquad& \hat{\Gamma}^{\text{(10D)}}_{m'}=I_4\otimes\gamma^{'}_{m'} \otimes \sigma_2.
\eea
We define
\beq
\hat{\Gamma}^{\text{(10D)}}_{*}=-\hat{\Gamma}^{\text{(10D)}}_{0}\ldots \hat{\Gamma}^{\text{(10D)}}_{9}=-I_4\otimes I_4\otimes\sigma_3.
\eeq
We can write $10$-dimensional spinors in this decomposition as
\beq
\Psi_{\text{(10D)}} =\psi \otimes \psi^{'} \otimes\left( \begin{matrix}
a \\
b
\end{matrix}\right),
\eeq
however, the type IIB chirality condition $\tfrac{1}{2}\left(1+\hat{\Gamma}^{\text{(10D)}}_{*}\right) \Psi_{\text{(10D)}}=0$ implies that
\beq
0=\tfrac{1}{2}\left(1+\hat{\Gamma}^{\text{(10D)}}_{*}\right) \Psi_{\text{(10D)}}=\psi\otimes \psi^{'}\otimes \left(\begin{matrix}
0\\
b
\end{matrix}\right)\qquad\rightarrow\qquad b=0.
\eeq
Our $10$-dimensional chiral spinor is then 
\beq
\Psi_{\alpha}^{\;\;i}=\psi_{\alpha}\otimes \psi^{'i}\otimes \left(\begin{matrix}
1\\
0
\end{matrix}\right),
\eeq 
where we reabsorbed the constant $a$ into the $4$-dimensional spinors. By doing this restriction to the right handed chiral subspace we can again define
\beq
\Gamma_{\hat{m}\hat{n}}=\hat{\Gamma}^{\text{(10D)}}_{\hat{m}\hat{n}}  \tfrac{1}{2}\left(1+\hat{\Gamma}^{\text{(10D)}}_{*}\right),
\eeq
and identify
\beq
\Gamma_{\tilde{m}\tilde{n}} = \gamma_{\tilde{m}\tilde{n}}\otimes I_4,\qquad \Gamma_{\tilde{m}'\tilde{n}'} = I_4\otimes \gamma^{'}_{\tilde{m}'\tilde{n}'},\qquad \Gamma_{\tilde{m}\tilde{n}'} = \gamma_{\tilde{m}}\otimes\gamma^{'}_{\tilde{n}'}.
\eeq

\section{The $SU(2,2|4)$ algebra in various forms}\label{SUMN}
The $SU(2,2|4)$ algebra is
\bea
[V_{\hat{\alpha}}^{\;\;\hat{\beta}},V_{\hat{\gamma}}^{\;\;\hat{\delta}}]=\delta_{\hat{\gamma}}^{\;\;\hat{\beta}}V_{\hat{\alpha}}^{\;\;\hat{\delta}}-\delta_{\hat{\alpha}}^{\;\;\hat{\delta}}V_{\hat{\gamma}}^{\;\;\hat{\beta}},&\qquad& [U_i^{\;\;j},U_k^{\;\;l}]= \delta_i^{\;\;l}U_{k}^{\;\;j} - \delta_{k}^{\;\;j}U_{i}^{\;\;l}\nonumber\\
\ [V_{\hat{\alpha}}^{\;\;\hat{\beta}},\mathcal{Q}_{\hat{\gamma}}^{\;\;i}] = \delta_{\hat{\gamma}}^{\;\;\hat{\beta}}\mathcal{Q}_{\hat{\alpha}}^{\;\;i}-\frac{1}{4}\delta_{\hat{\alpha}}^{\;\;\hat{\beta}}\mathcal{Q}_{\hat{\gamma}}^{\;\;i}, &\qquad& [U_i^{\;\;j},\mathcal{Q}_{\hat{\alpha}}^{\;\;k}] = \delta_i^{\;\;k}\mathcal{Q}_{\hat{\alpha}}^{\;\;j} -\frac{1}{4}\delta_{i}^{\;\;j}\mathcal{Q}_{\hat{\alpha}}^{k},\nonumber
\eea
\beq\label{SUMN::Algebra}
\{\mathcal{Q}_{\hat{\alpha}}^{\;\;i},\bar{\mathcal{Q}}_{j}^{\;\;\hat{\beta}} \} =\delta_{j}^{\;\;i}V_{\hat{\alpha}}^{\;\;\hat{\beta}} -\delta_{\hat{\alpha}}^{\;\;\hat{\beta}}U_{j}^{\;\;i},
\eeq
with all other commutators vanishing. The index $\hat{\alpha}$ runs over the values $1,\ldots 4$ in this section.\\
We will relate the fundamental representation of $SU(2,2)$ to the spinor of $SO(2,4)$. We rotate the generators $V_{\hat{\alpha}}^{\;\;\hat{\beta}}$ to $\hat{M}_{\hat{m}\hat{n}}$ by means of the $\hat{\gamma}_{\hat{a}\hat{b}}$ matrices given in (\ref{Clifford::24RHproj})
\beq
V_{\hat{\alpha}}^{\;\;\hat{\beta}}=\frac{1}{2}(\hat{\gamma}^{\hat{a}\hat{b}})_{\hat{\alpha}}^{\;\;\hat{\beta}}\hat{M}_{\hat{a}\hat{b}},\qquad \hat{M}_{\hat{a}\hat{b}}=-\frac{1}{4}(\hat{\gamma}_{\hat{a}\hat{b}})_{\hat{\alpha}}^{\;\;\hat{\beta}}V_{\hat{\beta}}^{\;\;\hat{\alpha}}.
\eeq
This is consistent by the fact that the $V_{\hat{\alpha}}^{\;\;\hat{\beta}}$ are traceless and (\ref{Clifford::ThatOneRel}). 
The other bosonic subalgebra, generated by $U$ will be considered as an internal group (an $R$-symmetry) for the remainder of this section. 
The conjugate spinor charge $\bar{\mathcal{Q}}_{i}^{\;\;\hat{\alpha}}$ is defined as the four-dimensional Dirac conjugate spinor,
\beq
\bar{\mathcal{Q}}_{i}^{\;\;\hat{\alpha}}=i[(\mathcal{Q}^i)^{\dagger}\gamma^0]^{\hat{\alpha}}.
\eeq
With this isomorphism realised, we have a superalgebra in terms of the generators
\beq
\mathbf{T}_{\Lambda}:\qquad\qquad \hat{M}_{\hat{a}\hat{b}},\qquad U_i^{\;\;j},\qquad \mathcal{Q}_{\hat{\alpha}}^{\;\;i}\qquad \text{and}\qquad \bar{\mathcal{Q}}_i^{\;\;\hat{\alpha}}.
\eeq
The super spacetime part of the algebra now gets the universal form
\bea\label{SUMN::SO2dalgebra}
[\hat{M}_{\hat{a}\hat{b}},\hat{M}_{\hat{c}\hat{d}}]&=& \hat{\eta}_{\hat{a}[\hat{c}}\hat{M}_{\hat{d}]\hat{b}}-\hat{\eta}_{\hat{b}[\hat{c}}\hat{M}_{\hat{d}]\hat{a}},\nonumber\\
\ [\hat{M}_{\hat{a}\hat{b}},\mathcal{Q}_{\hat{\alpha}}^{\;\;i}]&=& -\frac{1}{4}(\hat{\gamma}_{\hat{a}\hat{b}})_{\hat{\alpha}}^{\;\;\hat{\beta}}\mathcal{Q}_{\hat{\beta}}^{\;\;i},\nonumber\\
\ \{\mathcal{Q}_{\hat{\alpha}}^{\;\;i},\mathcal{Q}_{j}^{\;\;\hat{\beta}} \} &\sim& \delta_j^{\;\;i}(\hat{\gamma}^{\hat{a}\hat{b}})_{\hat{\alpha}}^{\;\;\hat{\beta}} \hat{M}_{\hat{a}\hat{b}} +\delta_{\hat{\alpha}}^{\;\;\hat{\beta}} U_i^{\;\;j},
\eea
and there is the internal part which involves the generators $U_i^{\;\;j}$, which also rotate the supercharges. The metric $\hat{\eta}=\text{diag}(-+++ + -)$ is the (2,4) flat metric and the indices $\hat{a}=\{0,1,2,3, S, T\}$ where $0$ and $T$ are timelike directions. Remark that we chose all generators in this formula to be dimensionless. In general we define a $\mathbf{G}$ valued object $A$ as
\beq
A=A^{\Lambda}\mathbf{T}_{\Lambda}.
\eeq
For the superalgebra above we have
\beq
A=\hat{A}^{\hat{a}\hat{b}}\hat{M}_{\hat{a}\hat{b}}+ \hat{A}_i^{\;\;j}U_j^{\;\;i} + \bar{\hat{A}}_i^{\;\;\hat{\alpha}}\mathcal{Q}_{\hat{\alpha}}^{\;\;i} +\bar{\mathcal{Q}}_i^{\;\;\hat{\alpha}}\hat{A}_{\hat{\alpha}}^{\;\;i},
\eeq
where these objects can be viewed as matrices. Note that $\bar{\mathcal{Q}}_i^{\;\;\hat{\alpha}}$ does not act on $\hat{A}_{\hat{\alpha}}^{\;\;i}$ in this notation.\\
We want to derive the generators of the $AdS$ algebra and the conformal algebra in their more familiar form. Starting from the generic form of the conformal superalgebra in the $SO(2,4)$ basis (\ref{SUMN::SO2dalgebra}), we will first decompose it into a form which is appropriate to the $AdS_{5}$ spacetime isometry algebra and then into a form which is appropriate for the conformal isometries in $4$ dimensions. We call these the $AdS$ decomposition and the conformal decomposition, respectively. We will also discuss how quantities in these decompositions are related.\\
\subsection{The AdS decomposition}\label{SUMN::AdSDecompositionSec}
The $AdS_{5}$ space is a $5$-dimensional manifold with structure group $SO(2,4)$, in order to obtain this from the algebra we split the generators into $SO(1,4)$ generators $\tilde{M}_{\tilde{m}\tilde{n}}$ and the remaining generators $\tilde{P}_{\tilde{m}}$, defined through
\beq
\tilde{P}_{\tilde{m}}=\frac{2}{R}\hat{M}_{\tilde{m}T},\qquad \tilde{M}_{\tilde{m}\tilde{n}}=\hat{M}_{\tilde{m}\tilde{n}}, \label{redefinedgenerators}
\eeq
where we have introduced the constant $R$, which has dimensions of a length to give the translations $\tilde{P}_{\tilde{m}}$ the canonical dimensions of $L^{-1}$. It will be associated with the radius of curvature of the $AdS$ space. The $S$-direction will be associated with the $AdS$ bulk direction.\\
The supercharges $\mathcal{Q}_{\hat{\alpha}}^{\;\;i}$ are rescaled to have dimensions $L^{-1/2}$,
\beq
\tilde{\mathcal{Q}}_{\hat{\alpha}}^{\;\;i}=R^{-1/2}\mathcal{Q}_{\hat{\alpha}}^{\;\;i}.
\eeq
This yields a superalgebra of the form
\bea
[\tilde{M}_{\tilde{a}\tilde{b}},\tilde{M}_{\tilde{c}\tilde{d}}]&=& \tilde{\eta}_{\tilde{a}[\tilde{c}}\tilde{M}_{\tilde{d}]\tilde{b}}-\tilde{\eta}_{\tilde{b}[\tilde{c}}\tilde{M}_{\tilde{d}]\tilde{a}},\nonumber\\
\ [\tilde{P}_{\tilde{a}},\tilde{M}_{\tilde{b}\tilde{c}}] &=& \tilde{\eta}_{\tilde{a}[\tilde{b}}\tilde{P}_{\tilde{c}]},\qquad [\tilde{P}_{\tilde{a}},\tilde{P}_{\tilde{b}}]=\frac{2}{R^2}\tilde{M}_{\tilde{a}\tilde{b}},\nonumber\\
\ [\tilde{M}_{\tilde{a}\tilde{b}},\tilde{\mathcal{Q}}_{\hat{\alpha}}^{\;\;i}]&=& -\frac{1}{4}(\hat{\gamma}_{\tilde{a}\tilde{b}})_{\hat{\alpha}}^{\;\;\hat{\beta}}\tilde{\mathcal{Q}}_{\hat{\beta}}^{\;\;i},\qquad [\tilde{P}_{\tilde{a}},\tilde{\mathcal{Q}}_{\hat{\alpha}}^{\;\;i}]=\frac{2}{R} (\hat{\gamma}_{\tilde{a}T})_{\hat{\alpha}}^{\;\;\hat{\beta}}\tilde{\mathcal{Q}}_{\hat{\beta}}^{\;\;i}\nonumber\\
\ \{\tilde{\mathcal{Q}}_{\hat{\alpha}}^{\;\;i},\tilde{\mathcal{Q}}_{j}^{\;\;\hat{\beta}} \} &\sim& \delta_j^{\;\;i}(\hat{\gamma}^{\tilde{a}T})_{\hat{\alpha}}^{\;\;\hat{\beta}} \tilde{P}_{\tilde{a}} + \frac{1}{R}\delta_j^{\;\;i}(\hat{\gamma}^{\tilde{a}\tilde{b}})_{\hat{\alpha}}^{\;\;\hat{\beta}} \tilde{M}_{\tilde{a}\tilde{b}} +\frac{1}{R}\delta_{\hat{\alpha}}^{\;\;\hat{\beta}} U_i^{\;\;j},
\eea
where $\tilde{\eta}=\text{diag}(-+++ +)$ is the flat metric with signature (1,4).\\
It is interesting to note that this algebra contains the dimensionful constant $R$, which can not be scaled away if we want the translations to have natural dimension of a mass.\\
For the $AdS$ superalgebra, we have the decomposition of a $\mathbf{G}$-valued object
\beq
A=\tilde{A}^{\tilde{a}}\tilde{P}_{\tilde{a}}+ \tilde{A}^{\tilde{a}\tilde{b}}\tilde{M}_{\tilde{a}\tilde{b}}+\tilde{A}_i^{\;\;j}U_j^{\;\;i} + \bar{\tilde{A}}_i^{\;\;\hat{\alpha}}\tilde{\mathcal{Q}}_{\hat{\alpha}}^{\;\;i} +\bar{\tilde{\mathcal{Q}}}_{i}^{\;\;\hat{\alpha}}\tilde{A}_{\hat{\alpha}}^{\;\;i}. 
\eeq
From this we infer that
\beq
\tilde{A}^{\tilde{a}}=R\hat{A}^{\tilde{a}T},\qquad \tilde{A}^{\tilde{a}\tilde{b}}=\hat{A}^{\tilde{a}\tilde{b}},\qquad \tilde{A}_i^{\;\;j}=\hat{A}_i^{\;\;j},\qquad \tilde{A}_{\hat{\alpha}}^{\;\;i} = R^{1/2} \hat{A}_{\hat{\alpha}}^{\;\;i}.
\eeq
\subsection{The Conformal decomposition}\label{ConformalDecomp}
We now turn to the conformal decomposition of the superalgebra. We obtain the conformal transformations (generators) in $4$ dimensions, i.e. translations ($P_a$), Lorentz transformations ($M_{ab}$), dilatations ($D$) and special conformal transformations ($K_a$), which also form the algebra SO(2,4), by
\bea
&P_a = \frac{2}{R} (\hat{M}_{aT}+\hat{M}_{aS}) = \tilde{P}_{a}+\frac{2}{R}\tilde{M}_{aS},&\qquad [P_a]=L^{-1},\nonumber\\
&M_{ab}=\hat{M}_{ab}=\tilde{M}_{ab},&\qquad [M_{ab}]=L^0,\nonumber\\
&D=2\hat{M}_{TS}=-R\tilde{P}_S, &\qquad [D]=L^0,\nonumber\\
&K_a=2R(\hat{M}_{aT}-\hat{M}_{aS})=R^2\tilde{P}_a -2R \tilde{M}_{aS}, &\qquad [K_a]=L,
\eea
where we indicated the natural length dimensions $L$. The tangent directions have now been split as $\hat{a}=\{\tilde{a},T\}=\{a,S,T\}$ and we introduce the $4$-dimensional Minkowski metric $\eta_{ab}=\text{diag}(-,+,+,+)$. It is natural to split the supercharge $\mathcal{Q}_{\hat{\alpha}}^{\;\;i}$ into two Lorentz supercharges, the supersymmetry $Q_{\alpha}^{\;\;i}$ and the conformal supersymmetry $S_{\alpha}^{\;\;i}$. One way to distinguish between the two is that they transform with opposite weight under the dilatations. Consider the commutator
\beq
[D,\mathcal{Q}_{\hat{\alpha}}^{\;\;i}]= -\frac{1}{2}\hat{\gamma}_{TS}\mathcal{Q}_{\hat{\alpha}}^{\;\;i}.
\eeq
Since $\hat{\gamma}_{TS}^2=1$ and $\text{Tr}\,\hat{\gamma}_{TS}=0$, we can define the projection operators
\beq
\mathcal{P}_{Q,S}=\frac{1}{2}(1 \pm \hat{\gamma}_{ST}).
\eeq
We note that $\hat{\gamma}_{ST}$ commutes with $\hat{\gamma}_{ab}$ and therefore preserves the $4$-dimensional Lorentz spinors as desired. This leads us to the following identification
\bea
\text{supersymmetry:} &\quad& Q_{\alpha}^{\;\;i} =\sqrt{2} R^{-1/2} \mathcal{P}_{Q}\mathcal{Q}_{\hat{\alpha}}^{\;\;i} = \sqrt{2} \mathcal{P}_{Q}\tilde{\mathcal{Q}}_{\hat{\alpha}}^{\;\;i} ,\quad [Q_{\alpha}^{\;\;i}]=L^{-1/2}\nonumber\\
\text{special supersymmetry:} &\quad& S_{\alpha}^{\;\;i} =\sqrt{2} R^{1/2} \mathcal{P}_{S}\mathcal{Q}_{\hat{\alpha}}^{\;\;i} = \sqrt{2} R \mathcal{P}_{S}\tilde{\mathcal{Q}}_{\hat{\alpha}}^{\;\;i} ,\quad [S_{\alpha}^{\;\;i}]=L^{1/2}
\eea
Bringing these decompositions into the algebra (\ref{SUMN::SO2dalgebra}) we obtain
\bea\label{SUMN::ConfDecAlgebra}
[M_{ab},M_{cd}]&=& \eta_{a[c}M_{d]b}-\eta_{b[c}M_{d]a},\nonumber\\
\ [P_a,M_{cd}]&=&\eta_{a[b}P_{c]},\qquad [K_a,M_{cd}]=\eta_{a[b}K_{c]},\nonumber\\
\ [D,P_a]&=&P_a, \qquad [D,K_a]=-K_a,\nonumber\\
\ [P_a,K_b]&=& = 2(\eta_{ab}D+2M_{ab}),\nonumber\\
\ [M_{ab},Q_{\alpha}^{\;\;i}] &=& -\frac{1}{4}(\hat{\gamma}_{ab}Q^i)_{\alpha},\qquad [M_{ab},S_{\alpha}^{\;\;i}] = -\frac{1}{4}(\hat{\gamma}_{ab}S^i)_{\alpha},\nonumber\\
\ [K_a,Q_{\alpha}^{\;\;i}]&=& -(\hat{\gamma}_{aT}S^{i})_{\alpha}, \qquad [P_a,S_{\alpha}^{\;\;i}]= -(\hat{\gamma}_{aT}Q^{i})_{\alpha},\nonumber\\
\ [D,Q_{\alpha}^{\;\;i}]&=& \frac{1}{2}Q_{\alpha}^{\;\;i},\qquad [D,S_{\alpha}^{\;\;i}] = -\frac{1}{2}S_{\alpha}^{\;\;i},\nonumber\\
\{Q_{\alpha}^{\;\;i},\bar{Q}_{j}^{\;\;\beta} \} &=& \delta_j^{\;\;i} (\hat{\gamma}^{aT})_{\alpha}^{\;\;\beta}P_a,\qquad \{S_{\alpha}^{\;\;i},\bar{S}_{j}^{\;\;\beta} \} = \delta_j^{\;\;i} (\hat{\gamma}^{aT})_{\alpha}^{\;\;\beta}K_a,\nonumber\\
\{Q_{\alpha}^{\;\;i},\bar{S}_{j}^{\;\;\beta} \} &=& \delta_j^{\;\;i} (\hat{\gamma}^{ab})_{\alpha}^{\;\;\beta}M_{ab} + \delta_j^{\;\;i}\delta_{\alpha}^{\;\;\beta}D - 2 \delta_{\alpha}^{\;\;\beta}U_j^{\;\;i}.
\eea
By having given appropriate dimensions to the generators, this algebra contains no dimensionful constants as opposed to the $AdS$-decomposition where it was unavoidable. \\
A superconformal object can be decomposed as follows
\beq
A= A_P^aP_a+A_M^{ab}M_{ab}+A_D D+A^a_KK_a +A_i^{\;\;j}U_j^{\;\;i} +\left(\bar{A}_{Qi}^{\;\;\;\;\alpha}Q_{\alpha}^{\;\;i} + \bar{A}_{Si}^{\;\;\;\;\alpha}S_{\alpha}^{\;\;i} +\text{h.c.}\right), 
\eeq
and this yields the following relations
\bea\label{SUMN::RelConfAndAdsDecomp}
A_P^a&=&\frac{R}{2}(\hat{A}^{aT}+\hat{A}^{aS}) = \frac{1}{2}\left(\tilde{A}^a+R\tilde{A}^{aS}\right),\nonumber\\
A_M^{ab}&=&\hat{A}^{ab}=\tilde{A}^{ab},\nonumber\\
A_D&=&\hat{A}^{TS}=-R^{-1}\tilde{A}^S,\nonumber\\
A_K^a&=&\frac{1}{2R}(\hat{A}^{aT}-\hat{A}^{aS})=\frac{1}{2R^2}(\tilde{A}^a-R\tilde{A}^{aS})\nonumber\\
A_i^{\;\;j}&=& \hat{A}_i^{\;\;j}=\tilde{A}_i^{\;\;j}\nonumber\\
\bar{A}_{Qi}^{\;\;\;\;\alpha} &=&\frac{R^{1/2}}{\sqrt{2}}\bar{\hat{A}}_{i}^{\;\;\;\;\hat{\beta}}(\mathcal{P}_Q)_{\hat{\beta}}^{\;\;\hat{\alpha}}=\frac{1}{\sqrt{2}}\bar{\tilde{A}}_{i}^{\;\;\;\;\hat{\beta}}(\mathcal{P}_Q)_{\hat{\beta}}^{\;\;\hat{\alpha}}\nonumber\\
\bar{A}_{Si}^{\;\;\;\;\alpha} &=&\frac{R^{-1/2}}{\sqrt{2}}\bar{\hat{A}}_{i}^{\;\;\;\;\hat{\beta}}(\mathcal{P}_S)_{\hat{\beta}}^{\;\;\hat{\alpha}}=\frac{1}{\sqrt{2}R}\bar{\tilde{A}}_{i}^{\;\;\;\;\hat{\beta}}(\mathcal{P}_S)_{\hat{\beta}}^{\;\;\hat{\alpha}}.
\eea
It is interesting to note that the translations and the special conformal transformations in the conformal decomposition \textit{mix} the $AdS$ translations and structure group rotations.\\
To conclude this section we give the $AdS$ objects in terms of the their conformal counterparts.
\bea\label{SUMN::RelAdSandConfDecomp}
\tilde{A}^a&=& A^{a}_P + R^{2} A_K^a,\nonumber\\
\tilde{A}^S&=& -R A_D, \nonumber\\
\tilde{A}^{aS}&=& R^{-1}A_P^a -RA_K^a,\nonumber\\
\tilde{A}^{ab} &=&A_M^{ab},\nonumber\\
\tilde{A}_{\hat{\alpha}}^{\;\;i} &=& \sqrt{2}\left(\begin{matrix}
A_{Q\alpha}^{\;\;\;\;i}\\
RA_{S\alpha}^{\;\;\;\;i}
\end{matrix}\right),
\eea
where just for notational reasons we have a basis in which $\hat{\gamma}_{ST}$ is diagonal.


\section{$AdS_5\times S^5$ as a coset space}\label{AdSxSCoset}
Our aim in this section is to construct the coset space $AdS_5 \times S^5$. First we consider $AdS_{5}$ as a coset space and then we discuss the $S^5$ coset space. We conclude this section with a discussion of an appropriate choice of fermionic coordinates for the coset superspace.
\subsection{$AdS_5$ as a coset space}
The $AdS_{5}$ space is the coset
\beq
AdS_{5}=\frac{SO(2,4)}{SO(1,4)}.
\eeq
The algebra to be considered is the bosonic part of the algebra in Appendix \ref{SUMN::AdSDecompositionSec} (ignoring the internal part). We choose horospherical coordinates,
\beq \label{AdSxS::AdSmetric}
ds^2=\rho^2 dx^2 +\left(\frac{R}{\rho}\right)^2d\rho^2,
\eeq
where the boundary is parametrized by $x^m$ and is at $\rho=\infty$. The coset representative for horospherical coordinates is given in the spinor representation of $SO(2,4)$. It can be derived from the supergravity Killing spinor \cite{Claus:1998yw} and can be written as
\beq\label{AdSCoset::Repr}
v(\tilde{x}^{\tilde{m}})=v_{\text{conf}}(x) \left(\rho^{-1/2}\frac{1}{2}\left(1-\hat{\gamma}_{ST}\right)+ \rho^{1/2}\frac{1}{2}\left(1+\hat{\gamma}_{ST}\right)\right),
\eeq
where $v_{\text{conf}}(x)$ is the coset representative of the $4$-dimensional conformal Minkowski space
\beq
v_{\text{conf}}(x) = 1+\frac{x^m}{R}\hat{\gamma}_{mT}\frac{1}{2}(1+\hat{\gamma}_{ST}).
\eeq
The flat $S$ direction is related to the bulk direction $\rho$ of $AdS_{5}$. Straightforward computation gives the Cartan forms (\ref{Cartan::DefCartan})
\beq
v^{-1}dv\equiv L^{\Lambda}T_{\Lambda} = e^{\tilde{m}}P_{\tilde{m}} + \omega^{\tilde{m}\tilde{n}}M_{\tilde{m}\tilde{n}},
\eeq
with non-vanishing components
\beq
e^m=dx^m \rho,\qquad e^{\rho}=d\rho \frac{R}{\rho},\qquad \omega^{m\rho}=dx^m\frac{\rho}{R}.
\eeq
The Killing fields $\Sigma_0$ (\ref{Cartan::Killingsuperfield}) are determined by an $\tilde{x}$-independent $SO(2,4)$ object,
\beq
\Upsilon=\tilde{a}^{\tilde{m}}\tilde{P}_{\tilde{m}}+ \tilde{\lambda}_M^{\tilde{m}\tilde{n}}\tilde{M}_{\tilde{m}\tilde{n}}.
\eeq
Using the $AdS$-decomposition
\beq
\tilde{P}_{\tilde{m}}=\frac{2}{R}\hat{M}_{\tilde{m}T},\qquad \tilde{M}_{\tilde{m}\tilde{n}}=\hat{M}_{\tilde{m}\tilde{n}},
\eeq
yields the $AdS_{5}$ Killing fields
\bea
\Sigma_0^m=\rho\xi^{m}(x) + \frac{R^2}{\rho}\lambda_{(K)}^m,&\qquad& \Sigma_0^\rho = -\Lambda_D(x)R,\nonumber\\
\Sigma_0^{m\rho}=\frac{\rho}{R}\xi^m(x)-\frac{R}{\rho}\Lambda^m_{(K)}, &\qquad& \Sigma^{mn}_0 = \Lambda_M^{mn}(x),
\eea
where the Killing fields have been written in terms of the conformal parameters. From these we get the Killing vectors and compensating (stability group) transformations. The Killing vectors (\ref{Cartan::SupIsometriesComps}) are given by 
\bea\label{AdSCoset::KillingVectors}
\xi^m &=& \hat{\Lambda}^{m}_{\;\;-} + \hat{\Lambda}^{mn} x_n + \hat{\Lambda}^{+}_{\;\;+} x^m + (x^2 \hat{\Lambda}^{m}_{\;\;+}-2x^m x_n \hat{\Lambda}^{n}_{\;\;+} ) + \frac{R^2}{\rho^2}\hat{\Lambda}^{m}_{\;\;+},\nonumber\\
\xi^{\rho}&=&-\hat{\Lambda}^{+}_{\;\;+} - 2x_{m}\hat{\Lambda}^{m}_{\;\;+}.
\eea
in terms of the independent parameters $\hat{\Lambda}^{m}_{\;\;-}$, $\hat{\Lambda}^{mn}$, $\hat{\Lambda}^{+}_{\;\;+}$ and $\hat{\Lambda}^{m}_{\;\;+}$, and the compensating transformations are
\beq
l^{mn}=\Lambda^{mn}_{(M)},\qquad l^{m\rho}=-2\frac{R}{\rho}\lambda^{m}_{(K)}.
\eeq
We can recast the Killing vectors into a form related to the conformal isometries on $x^m$. We define
\beq
a^{m}=\hat{\Lambda}^{m}_{\;\;-},\qquad \lambda^{mn}_{(M)}=\hat{\Lambda}^{mn},\qquad \lambda_D=\hat{\Lambda}^{+}_{\;\;+},\qquad \lambda_{(K)}^{m}=\hat{\Lambda}^{m}_{\;\;+},
\eeq
and obtain
\beq
\delta x^{m}=-\xi^{m}_{C}(x) - \frac{R^2}{\rho^2}\lambda_{(K)}^m,\qquad \delta \rho= \Lambda_D(x)\rho,
\eeq
with
\bea\label{AdSCoset::ConformalIsometries}
\xi^m_{C}(x)&=&a^m+\lambda^{mn}_{(M)} +\lambda_D x^m + (x^2\lambda_{(K)}^m - 2 x^m x\cdot \lambda_{(K)}),\nonumber\\
\Lambda^{mn}_{(M)}(x)&=& \lambda_{(M)}^{mn} - 4 x^{[m}\lambda_{(K)}^{n]},\nonumber\\
\Lambda_D(x)&=&\lambda_D - 2 \lambda_{(K)} \cdot x . 
\eea
Where $a^m$, $\lambda^{mn}_{(M)}$, $\lambda_D$ and $\lambda_{(K)}^{m}$ are the constant parameters of translations, Lorentz rotations, dilatations and special conformal transformations for the conformal space in four dimensions, spanned by the coordinates $x^{m}$. We have include the $C$ as a subscript for $\xi_{C}^{m}$ to stress that it is expressed in terms of the conformal parameters.
\subsection{$S^5$ as a coset space}\label{SCoset}
The sphere is the coset space
\beq
S^{5}=\frac{SO(6)}{SO(5)}.
\eeq
The algebra to be considered is the $SO(6)$ algebra
\beq
[\hat{M}'_{\hat{m}'\hat{n}'},\hat{M}'_{\hat{p}'\hat{q}'}]=\delta_{\hat{m}'[\hat{p}'}\hat{M}'_{\hat{q}']\hat{n}'} - \delta_{\hat{n}'[\hat{p}'}\hat{M}'_{\hat{q}']\hat{m}'},
\eeq
where in the sphere decomposition
\beq
\tilde{P}'{}_{m'}=\frac{2}{R}\hat{M}'_{m'S'},\qquad \tilde{M}'{}_{m'n'}=\hat{M}'_{m'n'}, \label{SCoset::redefinitionspheregenerators}
\eeq
with $m'$ the $5$ flat tangent directions of the sphere.\\
We will work in stereographic coordinates $z^{m'}$
\beq\label{AdSxS::Spheremetric} 
ds^2=\frac{4R^2}{(1+z^2)^2}dz^2,
\eeq
where $z^2=z^{m'}\eta_{m'n'} z^{n'}$.
The convenient coset representative for the sphere in these coordinates is
\beq\label{SCoset::SphereCosetRep}
u(z^{m'})=(1+z^2)^{-1/2}(1+z^{m'}\hat{\gamma}'_{m'S'}),
\eeq
given in the spinor representation
\beq
\hat{M}'_{\hat{m}'\hat{n}'}=\frac{1}{4}\hat{\gamma}'_{\hat{m}'\hat{n}'},
\eeq
where the matrices $\hat{\gamma}'_{\hat{m}'\hat{n}'}$ are elements of the $SO(6)$ Clifford algebra. Straightforward computation gives the Cartan forms (\ref{Cartan::DefCartan})
\beq
u^{-1} du = e^{m'}P_{m'} +\omega^{m'n'}M_{m'n'},
\eeq
with
\beq
e^{m'}=2R\frac{dz^{m'}}{1+z^2},\qquad \omega^{m'n'}=4\frac{z^{[m'}dz^{n']}}{1+z^2}.
\eeq
We introduce the rigid $SO(6)$-valued parameter $\Upsilon_S=\Lambda^{\hat{m}'\hat{n}'}\hat{M}'_{\hat{m}'\hat{n}'}$ and derive the Killing field  (\ref{Cartan::SupIsometriesComps}),
\bea
\Sigma^{m'}_0 &=& \frac{2R}{1+z^2}\left(\frac{1}{2}(1-z^2)\Lambda^{m'S'} + \Lambda^{m'n'}z_{n'}+z^{m'}z_{n'}\Lambda^{n'S'}\right),\nonumber\\
\Sigma^{m'n'}_0&=&\Lambda^{m'n'} + \frac{4}{1+z^2}\left(z^{[m'}\Lambda^{n']S'}+z^{[m'}\Lambda^{n']p'}z_{p'}\right),
\eea 
leading to the isometries
\beq
-\delta z^{m'}=\xi^{m'}=\frac{1}{2}(1-z^2)\Lambda'{}^{m'S'}+\Lambda'{}^{m'n'}z_{n'}+z^{m'}z_{n'}\Lambda'{}^{n'S}. \label{SCoset::sphereisometries}
\eeq
\subsection{$AdS_{5}\times S^5$ and adapted fermionic coordinates}
The bosonic space is of a direct product form
\beq
AdS_{5}\times S^{5}.
\eeq
The bosonic coset representative $g(X)$ then also takes the form of a direct product $g(X)=v\otimes u$, in terms of the bosonic representatives for $AdS$ and $S$ obtained before. We can enlarge this bosonic space to a superspace by the coset construction. We already derived the representatives for the bosonic subspaces. The only thing that is lacking is the fermionic coordinate choice, encoded in the matrix $e_{\dot{\alpha}}^{\;\;\alpha}$.\\
The conformal structure of the $AdS$ boundary and associated isometries is most apparent in the horospherical coordinates. The coordinates $x^m$, which parametrize the directions parallel to the boundary $\rho\rightarrow\infty$, can then be identified with the coordinates $x^m$ of the conformal Minkowski space. To continue this, we would like that half of the anticommuting coordinates of the $AdS\times S$ superspace can be identified with the $\theta$'s of the conformal superspace. This can be done by appropriately considering the relation between the $AdS$ and conformal decompositions. As a coordinate choice we will take
\bea\label{AdSSphere::FermCoordChoice}
\Theta&=&\bar{\Theta}_i \mathcal{Q}^i +\bar{\mathcal{Q}}_i \Theta^i\\
&=& (u^{-1})_{i}^{\;\;j}\bar{\theta}_j \rho^{1/2} Q^{i} +(u^{-1})_{i}^{\;\;j}\bar{\vartheta}_j \rho^{-1/2} S^{i} + \bar{Q}_i \rho^{1/2} \theta^{j} u_{j}^{\;\;i} +\bar{S}_i \rho^{-1/2} \vartheta^{j} u_{j}^{\;\;i}.\nonumber
\eea
The two coordinates $\{\theta,\vartheta\}$ together build up the anticommuting coordinate of the $AdS\times S$ superspace $\tilde{\theta}$ by
\beq
\tilde{\theta}^i_{\hat{\alpha}}=\sqrt{2}\left(
\begin{matrix}
  \theta^i_{\alpha} \\
  R \vartheta^i_{\alpha}
 \end{matrix}
 \right).
\eeq
We will call these coordinates the super-horospherical coordinates
\beq
Z^{M}=\{x^m,\rho, z^{m'}, \theta^{i},\vartheta^{i}\}.
\eeq
The parametrization for the fermionic symmetry parameter $\varepsilon$ will be 
\bea\label{AdSSphere::FermParam}
\bar{\varepsilon}_{i}^{\alpha}&=& \frac{1}{\sqrt{2}}(u^{-1})_{i}^{\;\;j}\left[R\rho^{-1/2}\bar{\eta}_j^{\hat{\beta}}\frac{1}{2}\left(1-\hat{\gamma}_{ST}\right)_{\hat{\beta}}^{\;\;\alpha} + \rho^{1/2}  \left(\bar{\epsilon}_{j}^{\;\;\hat{\beta}}+\bar{\eta}_{j}^{\;\;\hat{\gamma}}(\hat{\gamma}_{mT})_{\hat{\gamma}}^{\;\;\hat{\beta}}x^m\right)\frac{1}{2}\left(1+\hat{\gamma}_{ST}\right)_{\hat{\beta}}^{\;\;\alpha}\right]\nonumber\\
&=&\frac{1}{\sqrt{2}}(u^{-1})_{i}^{\;\;j}\left[R\rho^{-1/2}\bar{\eta}_j^{\beta}\frac{1}{2}\left(1+\gamma_5\right)_{\beta}^{\;\;\alpha} + \rho^{1/2}  \left(\bar{\epsilon}_{j}^{\;\;\beta}-\bar{\eta}_{j}^{\;\;\gamma}(\gamma_m)_{\gamma}^{\;\;\beta}x^m\right)\frac{1}{2}\left(1-\gamma_5\right)_{\beta}^{\;\;\alpha}\right].
\eea
This is determined by Killing spinor equation and its solution \cite{Claus:1998yw,Lu:1998nu}. We can make the same super-horospherical decomposition for the $\kappa$-symmetry parameter
\beq
\kappa_{+}^{\alpha}\mathbf{K}_{\alpha}= \left(\rho^{1/2}u_i^{j}\bar{Q}_j\kappa_{+Q}^i+\text{h.c.}\right) + \left(\rho^{-1/2}u_i^{j}\bar{S}_j\kappa_{+S}^i+\text{h.c.}\right),
\eeq
and the relationship between its irreducible components is modified with factors of $R$ such that $\kappa_{+Q}=R\beta_{-}\kappa_{+S}$ , or equivalently, $\kappa_{+S}=-\tfrac{1}{R}\beta_{+}\kappa_{+Q}$. 

\bibliographystyle{JHEP}
\bibliography{D3branes}

\end{document}